\pdfoutput=1
\documentclass[journal]{IEEEtran}
\ifCLASSINFOpdf
\else
\fi
\usepackage{amsmath,amsfonts}
\usepackage{subfloat}
\usepackage{mathtools}
\usepackage{graphicx}
\usepackage{float}
\usepackage{cuted}
\usepackage{stfloats}
\usepackage{caption}
\usepackage{xcolor}
\usepackage{capt-of}
\usepackage{booktabs}
\usepackage{threeparttable}
\usepackage[colorlinks,pdftex]{hyperref}
\usepackage[caption=false,font=normalsize,labelfont=sf,textfont=sf]{subfig}

\newcommand\blfootnote[1]{%
  \begingroup
  \renewcommand\thefootnote{}\footnote{#1}%
  \addtocounter{footnote}{-1}%
  \endgroup
}

\begin{document}
%
\title{DARTS: Diffusion Approximated Residual Time Sampling for Time-of-flight Rendering in Homogeneous Scattering Media}
%
%
%

\author{Qianyue He,
Dongyu Du,
Haitian Jiang,
Xin Jin*,
}
\vspace{-5em}
%
%

\markboth{arXiv preprint, Vol.1, No.1, June~2024}%
{Shell \MakeLowercase{\textit{et al.}}: Bare Demo of IEEEtran.cls for IEEE Journals}
%



\maketitle
\begin{strip}\centering
    \includegraphics[width=0.999\textwidth]{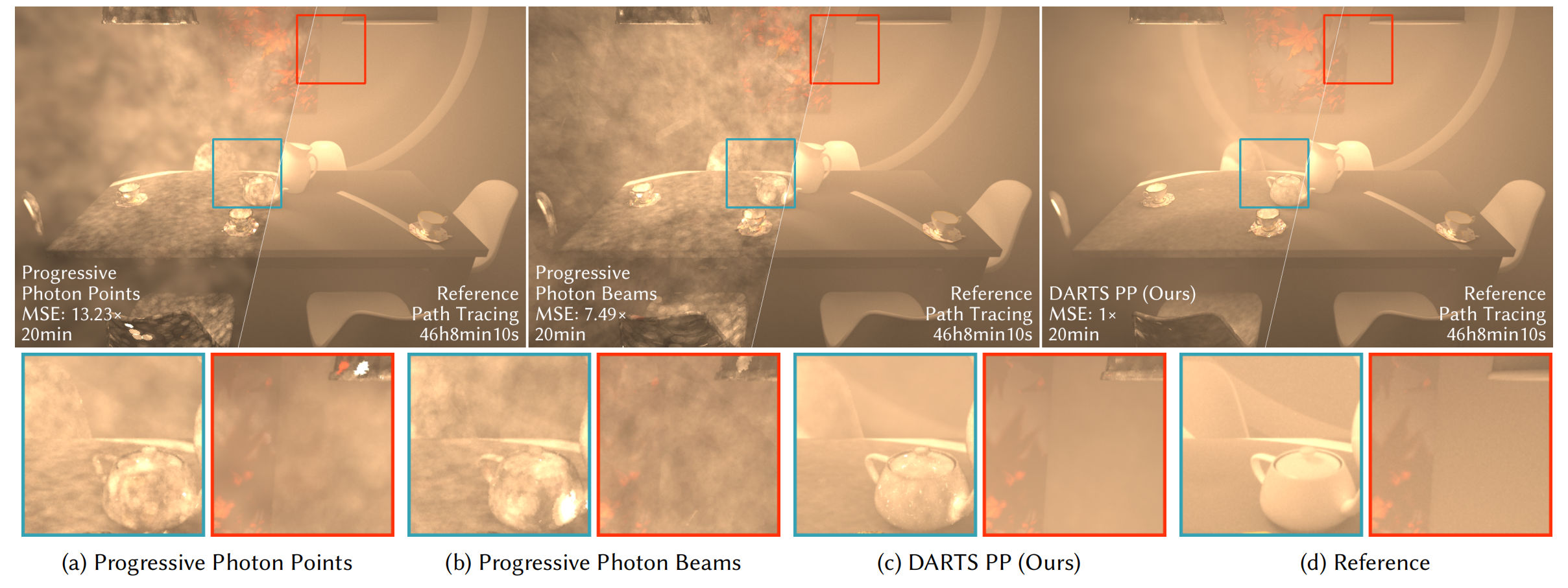}
    \captionof{figure}{Rendering of time-resolved transport using proposed DARTS in a scene with complex surface materials and homogeneous scattering media. DARTS integrates transient diffusion approximation into the path construction and adapts our elliptical sampling to provide path length control, enabling high quality time-of-light rendering and can be compatible with different existing frameworks. The example scene is illuminated by two non-synchronized pulse emitters with different start times of emission. Each image is rendered for 20 minutes. It can be seen that our sampling approach can greatly improve the SOTA photon based methods and provide lower overall MSE in the same rendering time. 
   }\label{fig:feature-graphic}
\end{strip}

\blfootnote{\hrule}
\blfootnote{Corresponding author. Email: jin.xin@sz.tsinghua.edu.cn}
\blfootnote{Qianyue He, Xin Jin, Haitian Jiang and Dongyu Du are with Shenzhen International Graduate School, Tsinghua University, Shenzhen, Guangdong 518000, China. }

\begin{abstract}
Time-of-flight (ToF) devices have greatly propelled the advancement of various multi-modal perception applications. However, achieving accurate rendering of time-resolved information remains a challenge, particularly in scenes involving complex geometries, diverse materials and participating media. Existing ToF rendering works have demonstrated notable results, yet they struggle with scenes involving scattering media and camera-warped settings. Other steady-state volumetric rendering methods exhibit significant bias or variance when directly applied to ToF rendering tasks. To address these challenges, we integrate transient diffusion theory into path construction and propose novel sampling methods for free-path distance and scattering direction, via resampled importance sampling and offline tabulation. An elliptical sampling method is further adapted to provide controllable vertex connection satisfying any required photon traversal time. In contrast to the existing temporal uniform sampling strategy, our method is the first to consider the contribution of transient radiance to importance-sample the full path, and thus enables improved temporal path construction under multiple scattering settings. The proposed method can be integrated into both path tracing and photon-based frameworks, delivering significant improvements in quality and efficiency with at least a 5x MSE reduction versus SOTA methods in equal rendering time.
\end{abstract}


\begin{IEEEkeywords}
transient rendering, time-gated cameras, participating media, modeling and simulation.
\end{IEEEkeywords}

%
\IEEEpeerreviewmaketitle

\section{Introduction}
The past decade has witnessed noteworthy advancements in time-of-flight (ToF) imaging methods which have the capability to capture transient responses of propagating photons utilizing ultra-fast sensors \cite{jarabo2017recent}. With the incorporation of temporal information, ToF imaging systems have revolutionized conventional imaging, making it possible to sense targets beyond the line of sight \cite{royo2023virtual} and operate well in extremely challenging environment, such as heavy fog \cite{du2022boundary}. Consequently, ToF devices propel the practical process of autonomous driving, robotic perception and scientific exploration, etc.

In tandem with ToF imaging techniques, ToF rendering provides an enhanced understanding of temporal data across various scene configurations, and thus contributes to the development of optical transmission theories and imaging algorithms, and the optimization of sensor systems \cite{marco2017deeptof, zhang2022first}. Moreover, ToF rendering plays a crucial role in generating extensive ToF datasets across diverse scenarios involving complex participating media, geometries and material properties, which are imperative for the advancement of data-driven methods \cite{attal2021torf}.

However, challenges reside in achieving efficient ToF rendering within participating media due to the complexity of temporal path construction. Despite substantial efforts directed towards steady-state scattering rendering \cite{herholz2019volume,lin2021fast}, the spatial sampling strategies fall short in generating temporal scattering paths. For transient rendering methods, certain methods achieve temporal uniformity in path samples \cite{jarabo2014framework, marco2019progressive}, but their accuracy is limited by the neglect of importance-sampling the contribution of transient radiance. Achieving effective importance sampling requires that major volumetric sampling processes, such as distance and direction sampling, incorporate transient radiance information. This requirement poses a challenge as the transient radiance information depends on global adjoint transport rather than just local transport functions. Additionally, the challenge persists in effectively imposing path time (length) constraints. The existing methods to impose path length constraints either prove not directly applicable for participating media \cite{pediredla2019ellipsoidal}, or inefficient for full transport and camera-warped settings \cite{liu2022temporally}, where scene-to-sensor transport time must be considered \cite{velten2013femto}. 

In this paper, we propose a diffusion approximated residual time sampling method (DARTS, for short), which provides full transient path construction and effective vertex connection within complex volumetric scenes and under camera-warped settings. To address the challenge of constructing effective temporal sampling paths in scattering media, instead of adopting uniform sampling in the time domain \cite{jarabo2014framework, marco2019progressive}, which overlooks the radiance contribution differences of different path samples, we perform importance sampling on the transient radiance by integrating the transient diffusion approximation (DA) into the rendering process. This approach allows us to obtain improved free-path distance and direction samples with enhanced overall radiance, leading to better convergence performance. To impose path time constraints, we extend the ellipsoidal connection method and further combine it with the proposed DA-based direction sampling to bypass challenges introduced by the reparameterization of ellipsoidal connection \cite{pediredla2019ellipsoidal} and avoid sampling inefficiency in camera-warped settings \cite{liu2022temporally} . The proposed method is inherently unbiased and introduces negligible extra computation and memory overhead. 

In particular, we make the following contributions:
\vspace{-0.5em}
\begin{itemize}
    \item We \textit{propose} a novel distance sampling method named DA distance sampling based on transient diffusion theory.
    
    \item We \textit{propose} a novel direction sampling method named elliptical DA direction sampling, by tabulating the transient DA values integrated in an equal-time ellipse and \textit{develop} corresponding multiple importance sampling (MIS) strategy.
    
    \item We \textit{extend} the ellipsoidal connection to volumetric rendering and \text{enable} its capability for importance sampling and direction reuse, to effectively control the path length.
    
    \item We \textit{demonstrate} the proposed method outperforms the existing method with at least a 5x MSE reduction and can be integrated as straightforward plug-ins for both path tracing and photon based frameworks.
\end{itemize}

The code for the proposed method in both \textit{pbrt-v3}\footnote{https://github.com/mmp/pbrt-v3} \cite{pharr2023physically} and \textit{Tungsten}\footnote{https://github.com/tunabrain/tungsten} \cite{Bitterli:2018:Tungsten} frameworks is provided in our supplementary materials.

\section{Related Work}
\subsection{Time-of-flight imaging devices}

ToF devices employ ultra-fast sensors to capture and count the photons received at different time points \cite{jarabo2017recent}. This time-resolved information is recorded to generate single time-gated image or sequences of temporal waveforms with high temporal resolution \cite{jarabo2012femto}, which can be applied in imaging through scattering media \cite{du2022boundary, wu2018adaptive} and obscurants \cite{9534479, kijima2021time}, non-line-of-sight imaging \cite{rapp2020seeing, faccio2020non, xin2019theory}, material estimation \cite{zickus2020fluorescence, shem2020towards} and improved depth sensing \cite{gruber2019gated2depth, walia2022gated2gated}.

\subsection{Time-of-flight rendering}
 \textit{Transient rendering}. It aims to simulate the temporal responses as a sequence of frames. Existing works can already guarantee physical correctness, from the basics of transient light transport for both forward \cite{jarabo2014framework} and differentiable \cite{yi2021differentiable} cases, to more complex vector light transport \cite{jarabo2018bidirectional} and in media with spatially various refraction index \cite{ament2014refractive}. However, the convergence for transient rendering is slow due to the lack of efficient sampling method \cite{bitterli2016}. Efforts have been made to importance-sample the uniform distribution of path lengths for better temporal density estimation and path reuse \cite{jarabo2014framework}. Subsequent methods \cite{marco2017transient, marco2019progressive} extend the photon beams method to transient state, but they introduce bias in exchange for reduced variance. The above methods heavily rely on temporal path reuse, making it challenging to be extended to time-gated rendering where temporal path reuse will lead to massive sample rejection. Also, existing works often disregard the importance sampling of transient radiance contributions during \textit{path construction} and therefore cannot guarantee a good approximation of the integrand for Monte Carlo integration. Other works propose different insights like using instant radiosity \cite{pan2019transient} for fast rendering, while the visual effects are constrained to diffuse scattering, or employ sampling methods biased towards receivers \cite{periyasamy2016importance, lima2011improved} and consider primarily the spatial distribution of radiance, therefore neglecting the temporal distribution. 

\textit{Time-gated rendering}. Researches focus on improving path connections to impose constraints on the path time (length). Pediredla et al. \cite{pediredla2019ellipsoidal} demonstrate that, given two vertices to be connected and a target path length, an ellipsoid can be defined and any connection with one intermediate vertex sampled on this ellipsoid has equal path length. Unfortunately, their parameterization cannot be directly extended to volumetric rendering, as volumetric sampling lacks intersecting polygons. Additionally, parameterizing the sampling space around the ellipsoid center prevents effective importance sampling and the reuse of sampled directions. Liu et al. \cite{liu2022temporally} extend the methods introduced by Deng et al. \cite{deng2019photon} to the time domain, illustrating that time-related sampling can be viewed as sampling from the sliced high-dimensional photon primitives. However, this method is not efficient under camera-warped settings \cite{velten2013femto} or with surface transport. These methods in general focus on improving \textit{path connection} for better path length control and do not yield better multiple scattering paths, which are crucial for high-order scattering scenarios and temporal importance sampling. 

\subsection{Rendering in homogeneous scattering media}
Various approaches have been developed \cite{novak2018monte} to improve the rendering quality in homogeneous scattering media. We will focus on two most widely adopted types of rendering methods, as our method is applicable to both.

\subsubsection{Density Estimation Based Methods}
We refer to these methods as \textit{photon based methods}. Stemming from photon mapping \cite{jensen1996global, jensen1998efficient}, this kind of two-pass biased estimator is later upgraded to progressive ones \cite{hachisuka2008progressive, hachisuka2009stochastic} . Jarosz et al. \cite{jarosz2011comprehensive} extend the point-point 3D kernel estimator to various point-beam and beam-beam estimators, which greatly alleviate the visual artifacts caused by photon sparsity and are later combined with bidirectional methods \cite{kvrivanek2014unifying}. Subsequent works \cite{bitterli2017beyond,deng2019photon} devise unbiased photon estimators by shrinking the kernel to a spatial delta-function to eliminate bias. Though photon based methods has temporal extensions \cite{marco2017transient, liu2022temporally}, they currently prove to be either ineffective for settings with reduced rendering time range, such as time-gated rendering, or inefficient for camera-warped settings. This inefficiency arises from the inability to importance sample transient radiance for constructing better paths, as well as the dependence on temporal path reuse \cite{jarabo2014framework}.
\vspace{-0.5em}\subsubsection{Monte Carlo Path Tracing}
Previous methods aim to analytically approximate one or more terms in the integrand during distance or direction sampling  \cite{kulla2012importance, georgiev2013joint}. More recent methods based on path guiding \cite{vorba2014line, herholz2016product} employ an online learning approach. Adjoint transport information is used to fit the local radiance distribution, which is later sampled to produce better ray directions. Herholz et al. \cite{herholz2019volume} further combine path guiding with zero variance random walk theory \cite{hoogenboom2008zero, ren2008gradient}, yielding a framework that can guide all scattering events. However, steady-state radiance lacks ToF information for path tracing, and extending path guiding directly to the time domain significantly increases training samples sparsity due to the curse of dimensionality.

Moreover, neither of the above methods adequately addresses the importance of more effective \textit{path construction}, which is crucial in suppressing variance for high-order scattering rendering. Therefore, we aim to establish a unified framework capable of both time-gated and transient rendering through optimizing both path construction procedures and path connection strategies. 

\section{Background}
Time-of-flight renderers lift the infinite speed of light assumption. Thus, the radiance transport theory and estimators should explicitly account for path time information. In the transient path integral framework proposed by Jarabo et al. \cite{jarabo2014framework}, the intensity $I$ of an image pixel is given by:

\begin{equation}\label{eqn:path-int-2}
\begin{aligned}
I = \int_{\Delta t_0}\int_{\Omega}
L_e(\mathbf{x}_e\rightarrow \mathbf{x}_{k-1}, t_e, \Delta t_e)
f(\overline{\mathbf{x}})\\
W(\mathbf{x}_{1}\rightarrow\mathbf{x}_0, \Vert \overline{\mathbf{x}}\Vert)
d\mu(\overline{\mathbf{x}})
d\Delta t_e,
\end{aligned}
\end{equation}
where $\overline{\mathbf{x}} = \mathbf{x}_0\mathbf{x}_1...\mathbf{x}_{k-1}\mathbf{x}_e$ is the simulated path with $k+1$ vertices; $\mathbf{x}_i\in \mathbb{R}^3$ denotes the position of path vertices; $\mathbf{x}_0$ and $\mathbf{x}_e$ denote sensor and emitter vertex, respectively; $d\mu(\cdot)$ is the Lebesgue measure; $\Omega$ denotes path space. $W(\mathbf{x}_{1}\rightarrow\mathbf{x}_0, \Vert \overline{\mathbf{x}}\Vert)$ is the response function of the sensor and is most relevant to temporal transport since it is the function of path optical length $\Vert \overline{\mathbf{x}}\Vert$. Since our work does not account for fluorescence and other microscopic scattering delays, factors that affect $\Vert \overline{\mathbf{x}}\Vert$ can only be the accumulated path length and emission duration $\Delta t_e$. $L_e(\mathbf{x}_k\rightarrow \mathbf{x}_{k-1}, t_e, \Delta t_e)$ is the emission term where $t_e$ denotes the start time of emission and $\Delta t_0$ denotes the integral space of $\Delta t_e$. Typically, for a simple pulse emitter with the start time of emission $t_e$ set to $0$, the simplified term $L_e$ can encompass most of the use cases. In the following, we will ignore the view dependence in $\mathbf{x}_{1}\rightarrow\mathbf{x}_{0}$ for simplicity. $f(\overline{\mathbf{x}})$ is path throughput term and takes the following form:
\begin{equation}\label{eqn:path-throughput}
f(\overline{\mathbf{x}}) = g(\overline{\mathbf{x}})G(\overline{\mathbf{x}}),
\end{equation}
$G(\overline{\mathbf{x}})$ is the throughput term consisting of path transmittance, measure conversion terms and visibility terms; $g(\overline{\mathbf{x}})$ denotes the local throughput function. In our work, both cases where $g(\overline{\mathbf{x}})$ is either the bidirectional scattering distribution function (BSDF) of the surface or phase function in the medium are considered. In scenes filled with homogeneous participating media with constant relative refraction index $\eta$, the transport time for a path $\overline{\mathbf{x}}$ with $k$ vertices can be simplified to Equation \eqref{eqn:path-len}:
\begin{equation}\label{eqn:path-len}
t(\overline{\mathbf{x}}) = \frac{\eta}{c}\Vert\overline{\mathbf{x}}\Vert = 
\frac{\eta}{c}\sum_{i=0}^{k-1}\Vert \mathbf{x}_i - \mathbf{x}_{i+1}\Vert,
\end{equation}
$c$ represents speed of light in vacuum. Note that path length $\Vert \overline{\mathbf{x}}\Vert$ has the same meaning as path time $t(\overline{\mathbf{x}})$ and only differs by a scalar scaling factor. Therefore, path length $\Vert \overline{\mathbf{x}}\Vert$ and path time $t(\overline{\mathbf{x}})$ will be used interchangeably, with a slight abuse of notation. 

\begin{figure}[htp]
\centering
\includegraphics[width=0.9\linewidth]{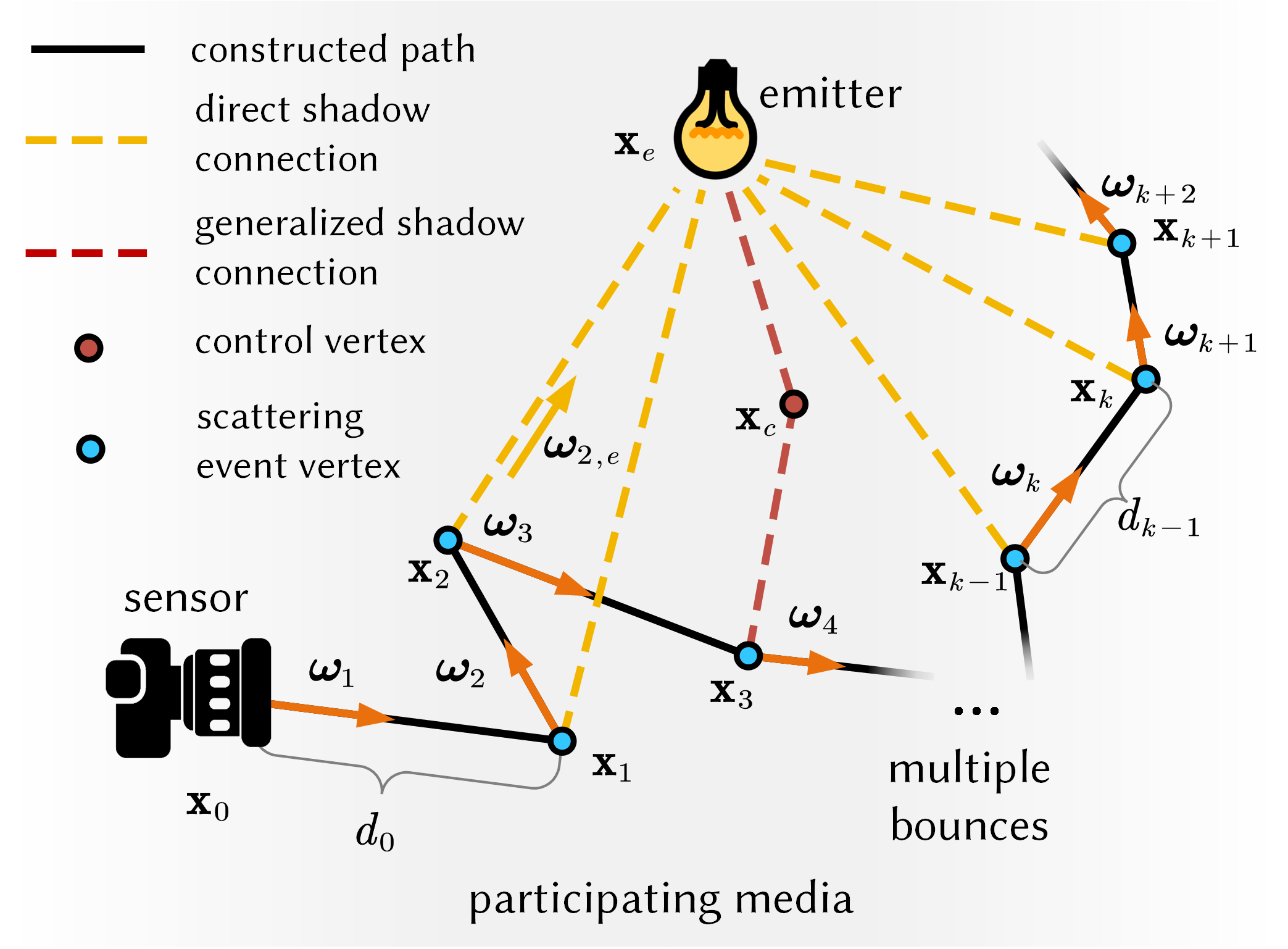}
\caption{Unidirectional path tracing in a scene filled with participating media. Random walk generates a path (solid black lines) with multiple vertices and all the vertices are connected to the emitter either through direct shadow connection (dashed yellow lines) or generalized shadow connection with control vertices (dashed red lines)\label{fig:full-process}}
\end{figure}

Equation \eqref{eqn:path-int-2} can be estimated numerically using Monte Carlo integration. A single path with multiple vertices originating from camera $\mathbf{x}_0$ is traced and each vertex is connected to the emitter, as shown in Figure \ref{fig:full-process}. The traced path, direct shadow connection and generalized shadow connection (connection with extra control vertices) are represented by solid black line, dashed yellow line and dashed red lines, respectively. The transient radiance estimator for the estimation of the integral defined by Equation \eqref{eqn:path-int-2} takes the form of the following \cite{pediredla2019ellipsoidal}:
\begin{equation}\label{eqn:full-estimator}
\tilde{I} = \frac{1}{N}\sum_{n=1}^N \frac{
W(\lVert \overline{\mathbf{x}_n}\rVert)
f(\overline{\mathbf{x}_n})
L_e
}
{p(\overline{\mathbf{x}_n})},
\end{equation}
where $N$ different paths $\overline{\mathbf{x}_n}, n=1,2,...,N$ are used to estimate the pixel intensity; $p(\overline{\mathbf{x}_n})$ denotes the sampling probability density function (PDF) to the joint distribution of $\overline{\mathbf{x}_n}$. It can be seen that the temporal response function $W(\lVert \overline{\mathbf{x}_n}\rVert)$ is the only different part from the steady-state estimator. 

The high variance in transient radiance estimators is caused by two primary factors: neglecting radiance contributions in importance sampling, which results in less effective multiple scattering paths, and the inability to effectively impose path length constraints. Detailed variance analysis can be found in our supplementary note (Section A.1). In Section 4, we address the first issue by using the diffusion approximation (DA) to enhance importance sampling for distance sampling. In Section 5, we tackle the second issue by combining the ellipsoidal connection methodology with DA, improving direction sampling and effectively imposing path length constraints.
 
\section{Diffusion Approximated Distance Sampling}

\subsection{Residual Time Estimator}
Different from existing methods that exhibit limited awareness of transient radiance distribution, our method aims to construct paths where all vertices are importance-sampled according to radiance contribution of the specific target time intervals. Denoting target time for the full path as $T_t$ and the time taken for photons travelling to the $k$-th vertex as elapsed time $T_{\text{e}, k}$, the \textit{residual time} $T_{\text{res}, k}$ for the $k$-th vertex can be defined as:
\begin{equation}\label{eqn:residual}
T_{\text{res}, k} = T_t -  T_{\text{e}, k} = T_t - \frac{\eta}{c}\sum_{i=0}^{k-1}\Vert \mathbf{x}_i - \mathbf{x}_{i+1}\Vert,
\end{equation}
We note that residual time $T_{\text{res}}$ is usually longer than the path time for direct connection, and thus, imposing path length constraints cannot be effectively achieved through naive direct connection. 

\begin{figure}[htp]
\centering
\includegraphics[width=0.98\linewidth]{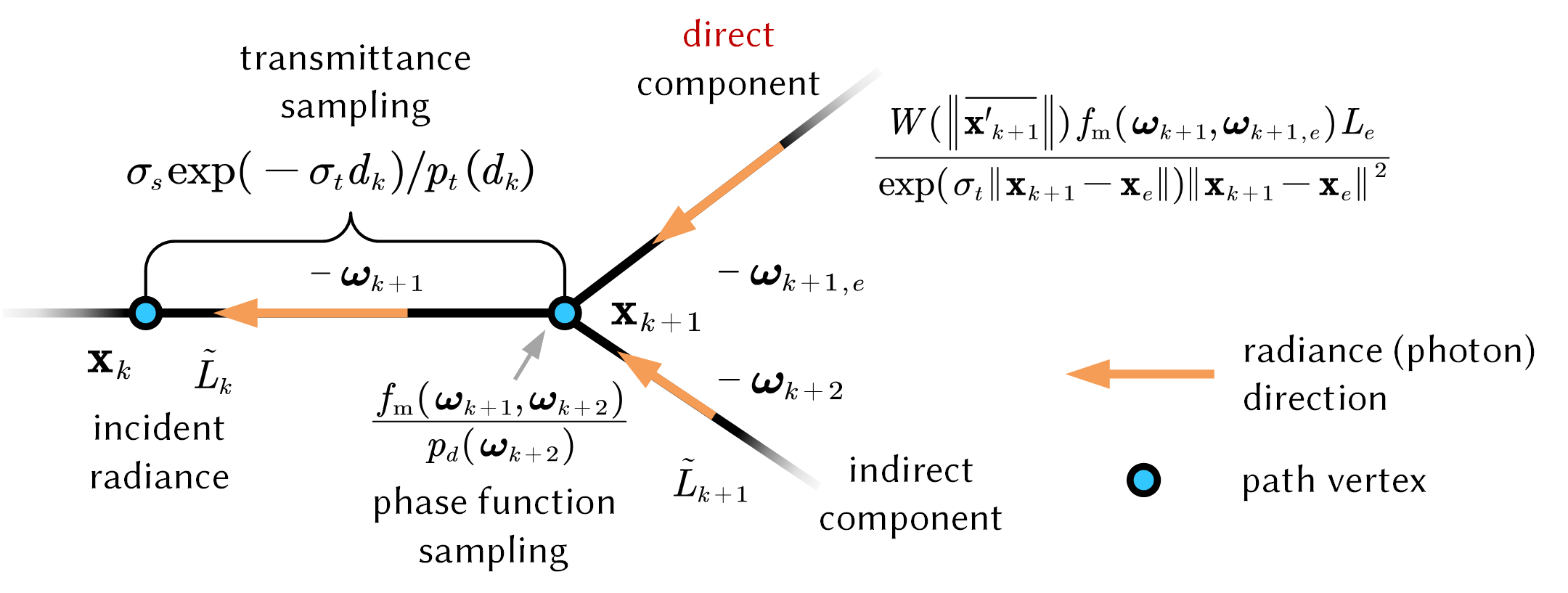}
\caption{Recursive decomposition of indirect radiance $\tilde{L}_k$ into the direct and indirect components. Through this decomposition, the estimator can actually be written as the summation of an infinite series of direct component at each vertex. This decomposition is easier to discuss, since it only depicts the local state transition.}\label{fig:recursive-form}
\end{figure}

To simplify the mathematical representation in Equation \eqref{eqn:full-estimator}, we formulate the transient radiance estimator in a recursive form. We decompose the incident radiance of vertex $\mathbf{x}_{k+1}$ into direct and indirect components, as depicted in Figure \ref{fig:recursive-form}. The combination of direct and indirect components at vertex $\mathbf{x}_{k+1}$ is the indirect radiance incident at $\mathbf{x}_{k}$ before exhibiting volumetric attenuation. The incident indirect radiance can then be defined with respect to the residual time:
\begin{equation}\begin{aligned}\label{eqn:our-est1}
\tilde{L}_k&(\mathbf{x}_{k}, -\pmb{\omega}_{k+1},T_{\text{res}, k}) = 
\biggl(
\underbrace{W(\Vert \overline{\mathbf{x}'_{k+1}}\Vert)
L_d(\mathbf{x}_{k+1}, T_{\text{res}, k+1}}_{\text{I}})
+\\
&\underbrace{\frac{f_\text{m}(\pmb{\omega}_{k}, \pmb{\omega}_{k+1})}
{p_d(\pmb{\omega}_{k+1})}
\tilde{L}_{k+1}}_{\text{II}}
\biggr) \frac{\sigma_s \exp(-\sigma_t d_k)}{p_t(d_k)},T_{\text{res}, k} = T_{\text{res}, k+1} + \frac{\eta}{c}d_k,
\end{aligned}\end{equation}
$\overline{\mathbf{x}'_{k+1}}$ denotes a path where the complete sensor-to-camera path is formed by connecting vertex $\mathbf{x}_{k+1}$ to $\mathbf{x}_e$, possibly via intermediate vertices. Part (I) in Equation \eqref{eqn:our-est1} is the exiting direct radiance at vertex $\mathbf{x}_{k+1}$, and here we use word \textit{direct} to denote connection paths produced by sampling control vertices before the sampled emitter, with a slight abuse of notation. Given the path length target $T_{t}$, valid sample must have $T_{\text{res}, k+1}$ as ToF to account for elapsed time of $\mathbf{x}_k$ and transport time for distance $d_k$. Likewise, for the indirect incident radiance $\tilde{L}_{k+1}$ in part (II), radiance sample with subsequent scattering should be able to satisfy the path length constraint, therefore it should have $T_{\text{res}, k+1}$ as ToF as well. As $\tilde{L}_{k+1}$ is sampled in scattering direction, the Monte Carlo term is applied to convert it to estimated exiting radiance. All components will be attenuated due to scattering and absorption within the transport distance $d_k$ produced by free-path distance sampling.  

Since both direct and indirect component have the same time $T_{\text{res}, k+1}$, Equation \eqref{eqn:our-est1} can be compressed to a unified residual time radiance representation that encompasses both direct and indirect components as $L'$, as shown in Figure \ref{fig:recursive-form}: 
\begin{equation}\begin{aligned}\label{eqn:our-est2}
\tilde{L}_k(\mathbf{x}_{k}, -\pmb{\omega}_{k+1},T_{\text{res}, k}) =\\ 
\frac{
\sigma_s \exp(-\sigma_t d_k)
L'(\mathbf{x}_{k+1}, -\pmb{\omega}_{k+1}, T_{\text{res}, k+1})
}
{p_t(d_k)},
\end{aligned}\end{equation}
$-\pmb{w}_{k+1}$ denotes the exiting direction at $\mathbf{x}_{k+1}$. To achieve low variance estimation, we need to find a sampling distribution whose PDF takes the following form:
\begin{equation}\label{eqn:zero-var1}
p_t(d_k) =
\frac{
\sigma_s \exp(-\sigma_t d_k)
L'(\mathbf{x}_{k+1}, -\pmb{\omega}_{k+1}, T_{\text{res}, k+1})
}
{Z},
\end{equation}
where $Z$ is the normalizing constant. The key point then comes down to sample the full paths according to the attenuated transient radiance $\exp(-\sigma_t d_k)L'(\mathbf{x}_{k+1}, -\pmb{\omega}_{k+1}, T_{\text{res}, k+1})$, in order to calculate the next vertex position $\mathbf{x}_{k+1}$. Equation \eqref{eqn:zero-var1} underscores the intuition behind using the unified residual time in Equation \eqref{eqn:our-est2}: sampling should be concentrated in areas where the combined direct and indirect components are substantial.

\vspace{-1.3em}
\subsection{Diffusion Approximated Sampling PDF }\label{sec:da-dist-samp}
To approximate adjoint transport solution $L'(\mathbf{x}_{k+1}, -\pmb{\omega}_{k+1},T_{\text{res}, k+1})$, we introduce the transient diffusion equation (DE) into the approximation of transient radiance. The radiant flux solution $\Phi(\mathbf{x}, t')$ of DE in an infinite homogeneous scattering medium satisifying $\sigma_s \gg \sigma_a$ \cite{contini1997photon} is given by :
\begin{equation}\begin{aligned}\label{eqn:da-3d}
\Phi(\mathbf{x}, t') = 
\frac{cH(t' - \tau)}
{\biggl[4\pi cD(t' - \tau)\biggr]^{3/2}}
&\exp(-\frac{\Vert \mathbf{x} - \mathbf{x}_e\Vert^2}{4cD(t' - \tau)} - \\ \sigma_ac(t'- \tau)),
\text{where }D = &\frac{1}{3(\sigma_a + \sigma_s (1 - g))},
\end{aligned}\end{equation}
$\tau$ is the time point when the delta wavefront is emitted; $\mathbf{x}$ and $t'$ denotes the position and time point to be evaluated, respectively; $g$ is the anisotropic coefficient of Henyey-Greenstein phase function, and is used to compute the reduced scattering coefficient $D$ \cite{lister2012optical}; $H(\cdot)$ is the Heaviside step function which prevents non-causality. This solution can be used to evaluate approximated radiance with specific residual time. Note that Equation $\eqref{eqn:da-3d}$ doesn't account for the direction information, therefore, the the directed radiance is approximated based on the radiant flux $\phi(\mathbf{x}, t')$ without direction information.

The above yields part of the distance sampling PDF, and transmittance is accounted for additionally. This would approximate the transient radiance emitted from the emitter, scattered in the medium (approximated transient radiance) and get attenuated from the sampled position to the ray starting position (transmittance). Therefore, the distance sampling PDF takes the following form: 
\begin{equation}\label{eqn:dist-samp}
p_t(d_{k}) = \frac{
\overbrace{\sigma_t\exp(-\sigma_t d_k)}^{\text{transmittance}}
\overbrace{\Phi(\mathbf{x}_{k} + \boldsymbol{\omega}_k d_k, T_{\text{res},k} - \frac{\eta}{c}d_k)}^{\text{flux approximated transient radiance}}
}{Z_k},
\end{equation}
$Z_k$ is the normalization constant. Substitute \eqref{eqn:dist-samp} into \eqref{eqn:our-est2}, we have:
\begin{equation}\label{eqn:our-simple-est1}
\tilde{L}_k(\mathbf{x}_{k}, -\pmb{\omega}_{k+1},T_{\text{res}, k}) = 
\underbrace{\frac{
	L'(\mathbf{x}_{k+1}, -\pmb{\omega}_{k+1}, T_{\text{res}, k+1})
}
{
	\Phi(\mathbf{x}_{k} + \pmb{\omega}_k d_k, T_{\text{res},k} - \frac{\eta}{c}d_k)
}}_{\text{variance inducing}}
\frac{\sigma_s Z_k}{\sigma_t},
\end{equation}
\begin{figure*}[htp]
\centering
\subfloat[Illustration for DA distance sampling]{\includegraphics[width=0.6\textwidth]{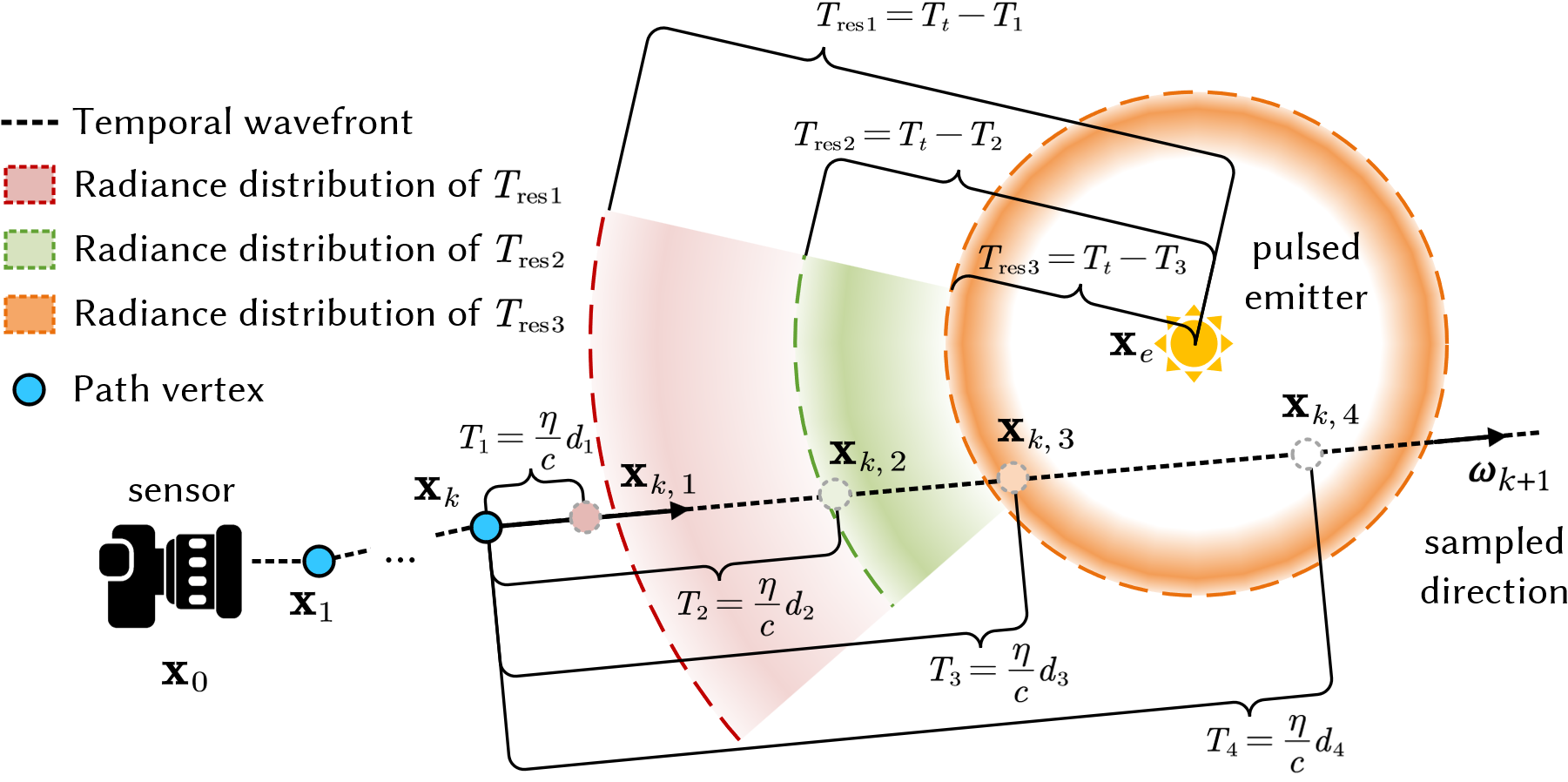}}
\subfloat[Illustration for 1D radiance distribution]{\includegraphics[width=0.4\textwidth]{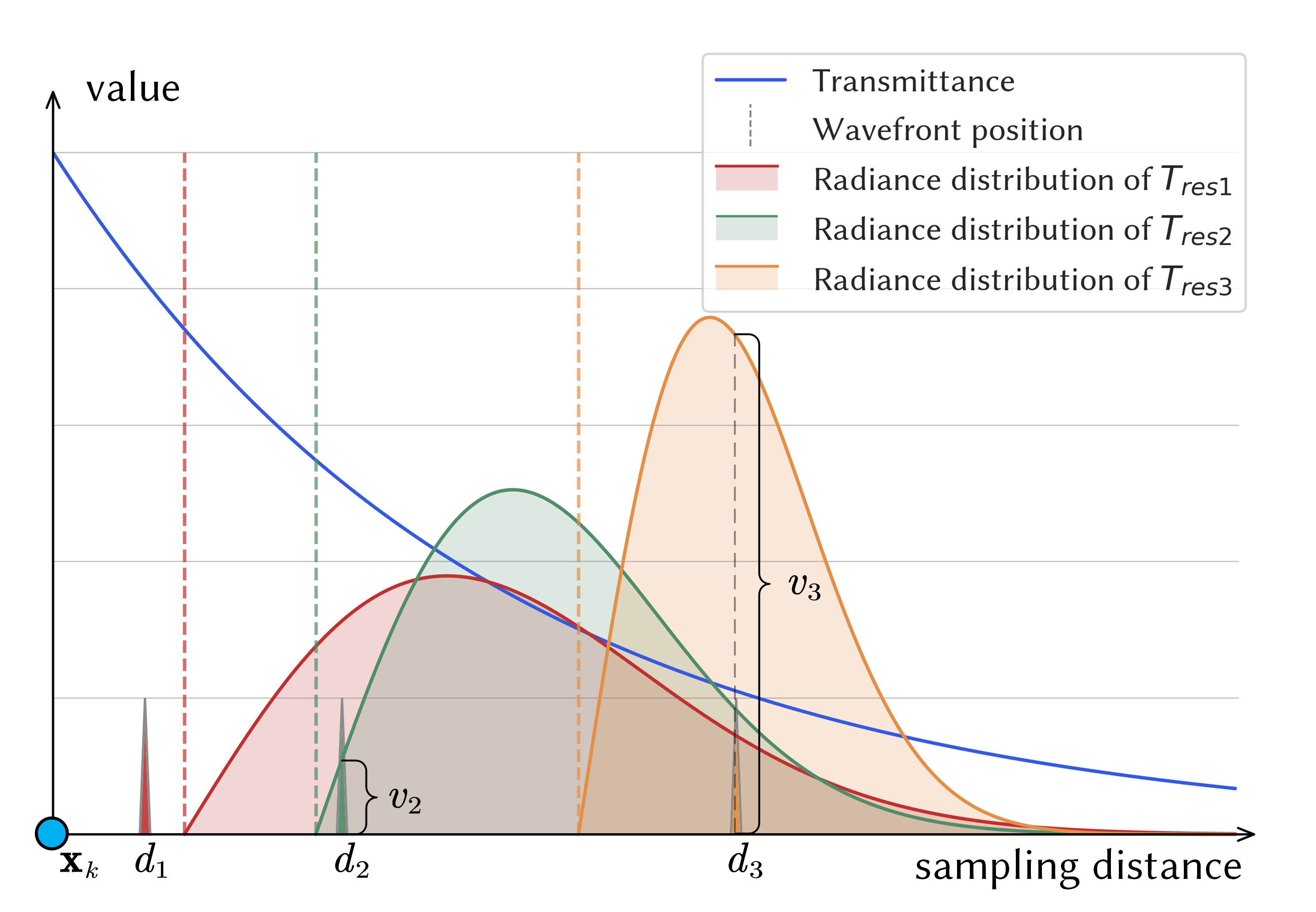}}
\caption{Illustration of DA distance sampling. Samples are drawn with statistically higher contribution according to product of transmittance and approximated transient radiance: 4 candidate samples $d_i$ are depicted in (a). $d_1$ and $d_4$ are invalid. Though $d_2$ has higher transmittance, the incident contribution on vertex $\mathbf{x}_k$ is lower than that of $d_3$, due having lower estimated radiance. Note that $d_4$ is not presented in (b), since it is invalid due to non-causality. }
\label{fig:figure4}
\end{figure*}
It can be seen that the denominator of the variance inducing term in Equation \eqref{eqn:our-simple-est1} is the approximation to the directionally integrated numerator, given the physical meaning of radiant flux. Also, it has been proved that when $\sigma_s \gg \sigma_a$ \cite{contini1997photon}, DA can accurately describe the radiant flux distribution and the radiance will be distributed uniformly in terms of direction, thus enabling the flux approximation for radiance. In this case, the variance inducing part can be regarded as a constant. Therefore, the proposed sampling approach achieves much lower variance compared to the existing methods while simultaneously incorporating path time information.

\subsection{Sample Generation}\label{sec:sec-4-3}
Unfortunately, the PDF given by Equation $\eqref{eqn:dist-samp}$ can not be analytically integrated due to its mathematical complexity, which makes analytical inverse sampling infeasible. Therefore, we generate samples according to Equation $\eqref{eqn:dist-samp}$ via resampled importance sampling (RIS) \cite{talbot2005importance}. The sample generation process is as follows:

\subsubsection{Scattering event sampling} Since PDF in Equation $\eqref{eqn:dist-samp}$ is actually conditioned on scattering events, we start by sampling the scattering events using exponential sampling. For a given ray, if the maximum travelling distance before hitting an opaque surface is $d_m$, the probability for medium scattering event can be given by:
\begin{equation}
p_m \coloneqq P(d_k < d_m) = \int_{0}^{d_m}\sigma_t \exp(-\sigma_t t)dt = 1 - \exp(-\sigma_t d_m),
\end{equation}
The sampled scattering event can then be defined by a Bernoulli distribution $ \text{Bern} (p_m)$ and can be decided by sampling from this distribution. DA distance sampling will be used if the current scattering event is sampled as a medium event.

\subsubsection{Candidate sample generation} For medium scattering events, distance samples can be obtained through Equation $\eqref{eqn:dist-samp}$ via RIS. We use truncated exponential distribution as our candidate distribution, since the local radiance transport is largely influenced by transmittance. Although other sampling methods such as equiangular sampling \cite{kulla2012importance} have been considered as proposals, they are found to be less effective (See supplementary note Figure VII). 
The candidate sampling distribution in RIS takes the following form:
\begin{align}
&p_{\text{candidate}}(d_k|d_k < d_m) = 
\frac{\sigma_t\exp(-\sigma_t d_k)}
{p_c}, d_k \in [0, d_m),\\
&d_k = \log\left(1 - p_c\epsilon\right) / \sigma_t, \epsilon \sim U[0, 1),p_c = 1 - \exp(-\sigma_t d_m)
\end{align}

\subsubsection{Evaluating transient DE and transmittance} Then, the sampled candidates are used to evaluate the RIS weights. After sampling according to the RIS weights, we can get the following estimator for the transmittance:
\begin{equation}\begin{aligned}
&\tilde{I}(d_k) = \frac{\exp(-\sigma_t d_k)}{\hat{p}(d_k)N_{\text{RIS}}}\sum_{i=1}^{N_{\text{RIS}}} w(d_k), w(d_k) = \frac{\hat{p}(d_k)}{p_{\text{candidate}}(d_k)},
\end{aligned}\end{equation}
$\hat{p}(d_k)$ is the denominator of Equation \ref{eqn:dist-samp}. The number of candidates $N_{\text{RIS}}$ is chosen to be the power of 2, and we use 8 to balance SIMD vectorization efficiency and output quality through out our implementation. To further enhance resampling efficiency, we propose to randomly sample one row from a pre-computed table (32 by 8) of Sobol sequence, then randomly offset each element in the sampled row by a uniformly distributed factor $\epsilon_0$ before the modulo operation into $[0, 1)$ range. Pre-computed Sobol sequence prevents the clustering of candidate samples. It also maintains good randomness and low correlation, while reducing the heavy computational load induced by Sobol sampler state updates. The RIS procedure employed eliminates the resolution-quality trade-off of tabulated sampling while remaining memory-efficient and can be further improved by reservoir resampling \cite{bitterli2020spatiotemporal}.
\begin{figure*}[h!tb]
\subfloat[EDA direction sampling tabulation]{\includegraphics[width=0.48\textwidth]{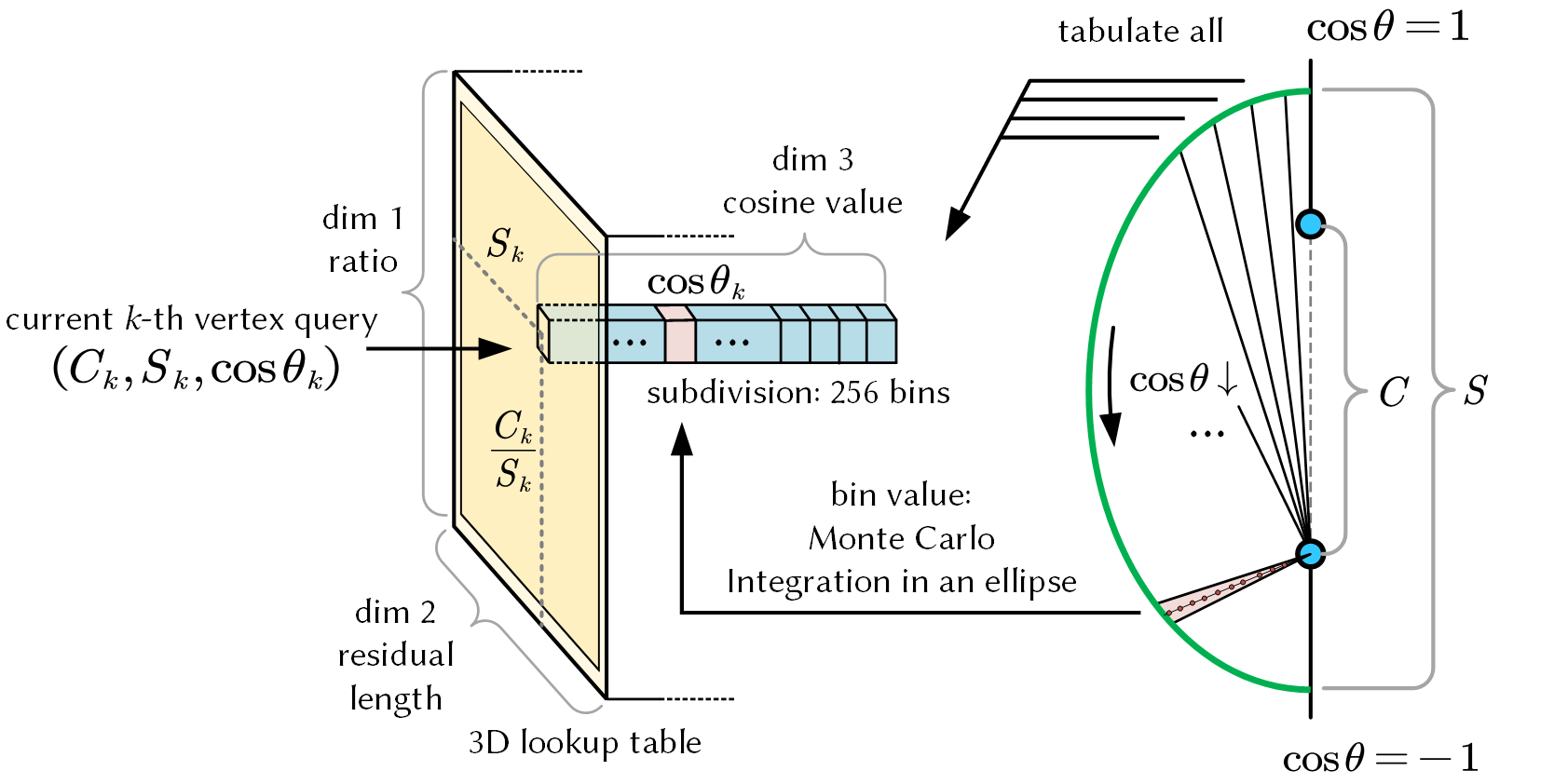}}
\subfloat[Two cases for elliptical sampling]{\includegraphics[width=0.48\textwidth]{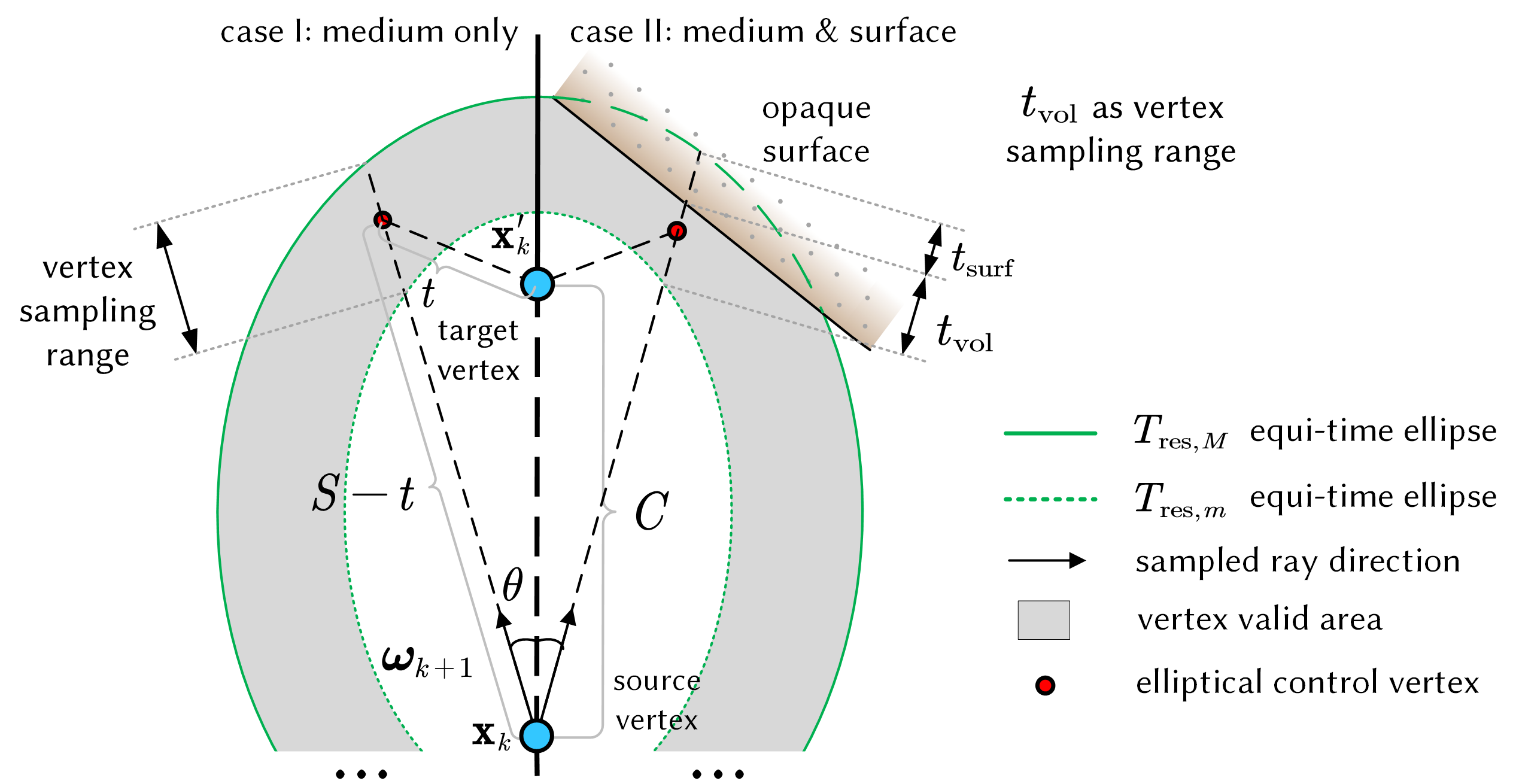}}
\caption{Two major sampling procedures in EDA sampling. The offline tabulation yields a 3D table for any vertex to query. This table is used for inverse transform sampling. Each value in the table is estimated via Monte Carlo integration on GPU (a).  Two possible elliptical sampling cases: the equi-time ellipse does not intersect any surfaces (case I, left half); surface is encountered within the sampling range (case II, right half) (b)}
\label{fig:figure6}
\end{figure*}

We provide an intuitive illustration in Figure \ref{fig:figure4} with 4 candidate distance samples $d_i, i=1,2,3,4$. Note that $\mathbf{x}_{k, i}$ in the figure are all candidate samples, the actual vertex for the $k$-th scattering event will be resampled from them. Since the sum of $T_{\text{res}}$ and the sampling time $T_i$ should be the given target time $T_t$, a longer sampling distance results in less propagation time (see $\mathbf{x}_{k,3}$ and the \textcolor[HTML]{EC8F5E}{tangerine wavefront} of $T_{\text{res}3}$), and vice versa (see $\mathbf{x}_{k,1}$ and the \textcolor[HTML]{BE3144}{scarlet wavefront} of $T_{\text{res}1}$). The sample with a higher product value of transmittance and approximated transient radiance is more likely to be resampled. Note that $d_1$ and $d_4$ are invalid since $\mathbf{x}_{k, 1}$ is out of the wavefront range and $\mathbf{x}_{k, 4}$ results in negative residual time. The weights of both samples are set to zero. In rare cases, if all candidates are invalid, this sampling approach will degrade to naive exponential sampling. The detection of invalid samples will be discussed in Section \ref{sec:ell-path-con}.

\section{Elliptical Diffusion Approximated Sampling}

In addition to distance sampling, efficient transient rendering requires effective direction sampling and path length control strategies. In this section, we propose a novel direction sampling method that incorporates elliptical path length control and diffusion approximation. We further employ our ellipsoidal connection \cite{pediredla2019ellipsoidal} extension, known as \textit{elliptical sampling}, within volumetric media for path length control. The integration of these two complementary modules is referred to as \textit{elliptical diffusion approximated} (EDA) sampling.

\subsection{EDA Direction Sampling}

The radiant flux in DA employs first-order spherical harmonic approximation and integrates direction away. We then opt for approximation that can capture direction information of transient radiance. 

Before starting a new path sample, a time interval will be sampled (deterministic for time-gated rendering) as the interval for target path time, denoted as $[T_{t, m}, T_{t, M})$. The minimum and maximum residual time range are given by $T_{\text{res}, m} = T_{t, m} - T_{e, k}$ and  $T_{\text{res}, M} = T_{t, M} - T_{e, k}$, respectively, where $T_{e, k}$ is the elapsed time of the path. Importance sampling requires direction sampling to obtain scattering direction that generates radiance samples (1) with high contribution and (2) with the ToF ranging $[T_{\text{res}, m}, T_{\text{res}, M})$. Since the upper bound of the residual time at $k$-th vertex is $T_{\text{res}, M}$, for any given direction, the next scattering vertex must reside in an ellipsoid defined by the current vertex, target emitter vertex and residual time, as the next path vertex outside of this ellipsoid will result in longer paths than required. Therefore, we approximate the incident radiance within the ellipse in direction $-\pmb{\omega}$ by the following equation:
\begin{align}
L_i(\mathbf{x}_k, -\pmb{\omega}, T) &= \sigma_s\int_{0}^{t_M} \exp(-\sigma_t t) L_o(\mathbf{x}_k + \pmb{\omega}t, -\pmb{\omega}, T - \frac{\eta}{c}t) dt \label{eqn:dir-exact}\\
&\approx \sigma_s\int_{0}^{t_M} \exp(-\sigma_t t) \Phi(\mathbf{x}_{k} + \pmb{\omega}_k t, T - \frac{\eta}{c}t)dt \label{eqn:dir-approx}
\end{align}
As shown in Equation \eqref{eqn:dir-approx}, $L_o$ can also be approximated by radiant flux introduced in Equation \eqref{eqn:da-3d}. Therefore, if the approximation is sufficiently accurate, we can effectively obtain the volumetric incident radiance from any given direction and for any specific ToF using the convolution of transmittance and DA within an ellipsoid. $t_M$ denotes the polar distance given $cT / \eta$ as the path length, which can be calculated deterministically:
\begin{equation}\label{eqn:polar-dist}
t_M(\cos\theta) = \frac{S^2 - C^2}{2S - 2C\cos\theta}, S = \frac{cT}{\mu},
\end{equation}
where $C$ is the focal distance between two vertices being connected; $S$ denotes the length sum of the connection paths; $\theta$ is the angle between the sampled direction $\pmb{\omega}$ and the major axis vector of the ellipsoid. 

Unfortunately, the integral in Equation \eqref{eqn:dir-approx} has no analytical form. In order to draw direction samples from approximated incident radiance, we tabulate Equation \eqref{eqn:dir-approx} by a 3D table. As shown in Figure \ref{fig:figure6} (a), the first two dimensions, $C / S$ and $S$ are tabulated for conditioning, since $C$ (distance to target) and $S$ (residual time) determine the shape of the ellipsoid, and are known for the given vertex. The resulting direction sampling PDF is conditioned on the two parameters that will be online-queried. We use $C/S$ instead of $C$ for the first dimension since the former is bounded in $[0, 1)$. The third dimension is the angular dimension used for inverse transform sampling. Note that $\cos\theta$ sampled by our method is the cosine value for the angle between sampled direction and ellipsoid major axis. Interval $[-1, 1]$ is uniformly subdivided into 256 bins, and for each bin, we evaluate Equation \eqref{eqn:dir-approx} via Monte Carlo integration. The $\phi$ angle is considered isotropic and sampled uniformly in $[-\pi, \pi)$ on the tangent plane defined by the major axis. Therefore, tabulation is evaluated in a 2D ellipse instead of a 3D ellipsoid. Thus we refer to this method as \textit{elliptical} DA instead of \textit{ellipsoidal}.  

The tabulation is calculated offline and parallel computation is straightforward. Our offline tabulation only takes around 5 seconds on a single consumer-end GPU and therefore the time overhead is negligible (refer to Section B.3 in supplementary note for more detail). Since we choose to sample $\cos\theta$ and $\phi$, the measure of the sampling PDF is mathematically equivalent to solid angle measure, and can thus be directly combined with phase function sampling via one-sample-model MIS \cite{veach1998robust} with balance heuristic. In our implementation, we choose the following adaptive parameter to decide when to use the proposed sampling method:
\begin{equation}
\gamma = \frac{C/S}{C/S + \alpha}\in [0, 1], \alpha \geq 0,
\end{equation}
where $\alpha$ is a parameter that controls the preference over proposed sampling, and we usually choose $0.5$ for experiment. $\gamma$ defines the probability of choosing the proposed method, and it is adaptive to the shape of the ellipse: when the ellipse resembles a circle (as $C/S$ is close to 0, when bounce count is low), phase function sampling is preferred; as the ellipse flattens after simulating multiple scattering, the proposed sampling is preferred.

To our knowledge, existing work does not importance-sample transient radiance contributions for paths of a given length. In contrast, our DA distance sampling and EDA direction sampling methods effectively address this challenge.
 
\subsection{Elliptical Sampling}\label{sec:ell-path-con}

Ordinary shadow connection fails to impose path length constraints for given vertices. We instead adapt the idea of ellipsoidal connection \cite{pediredla2019ellipsoidal} and sample a control vertex in participating media for path length control. However, the core reparameterization proposed by Pediredla et al. \cite{pediredla2019ellipsoidal} parameterizes the polygon around the center of the (projected) ellipsoid and therefore cannot be directly applied for vertex sampling in participating media, where there is no polygon to intersect. Additionally, parameterizing around the ellipsoid center prevents importance sampling and the reuse of sampled direction. We therefore introduce another parameterization to achieve efficient sampling in scattering media and further enable both importance sampling and direction reuse.

We choose to parameterize the sampling space around the current path vertex (focal point) in the polar coordinate system, meaning that the control vertex is parameterized by a direction $\pmb{\omega}$ from the current vertex $\mathbf{x}_k$ and a polar distance $t$. The control vertex can be sampled from a 2D elliptical ring in isotropic scattering media, given the residual time range $[T_{\text{res}, m}, T_{\text{res}, M})$, as shown in Figure \ref{fig:figure6} (b). We begin by redefining the PDF of elliptical vertex position, which is given by:
\begin{equation}
p(\mathbf{x}_{\text{ell}}) = p(\mathbf{x}_k + \pmb{\omega}t) \rightarrow p(\pmb{\omega}, t | \mathbf{x}_k) = p(t|\pmb{\omega}, \mathbf{x}_k)p(\pmb{\omega}| \mathbf{x}_k),
\end{equation}
$\mathbf{x}_{\text{ell}}\in \mathbb{R}^{3}$ is the position of the control vertex (red dot in Figure \ref{fig:figure6}). The distribution of $\mathbf{x}_{\text{ell}}$ can be represented by the joint distribution of the sampled direction $\pmb{\omega}$ and polar distance $t$, given one of foci located at $\mathbf{x}_k$. The representation is therefore decomposed into the product of two conditional probabilities defined with a 1D distance measure and a solid angle measure, lowering the sampling difficulty. Given the current vertex position $\mathbf{x}_k$, connection direction $\pmb{\omega}$ will be sampled first. Note that we reuse the scattering direction from the current vertex. That is, the next path vertex $\mathbf{x}_{k+1}$, $\mathbf{x}_{\text{ell}}$ and the current vertex $\mathbf{x}_{k}$ are on the same line. The ray-scene intersection results can thus be reused and can save much rendering time in high triangle-count scenes.

Since $\pmb{\omega}$ is reused and generated by EDA sampling, our discussion will focus on the sampling of $t$. The sampling of distance $t$ entails two primary scenarios, depending on whether the surface interaction is considered. In the case where the elliptical ring does not intersect any surface in the given direction, the $t$ sample depicted in the left half (case I) in Figure \ref{fig:figure6} (b), can be generated by the following equation:
\begin{equation}\label{eqn:ell-samp}
\begin{aligned}
t = \frac{S^2 - C^2}{2S - 2C\cos\theta}, S \sim P[S_m, S_M),\\ S_m = \frac{cT_{\text{res}, m}}{\eta}, S_M = \frac{cT_{\text{res}, M}}{\eta}
\end{aligned}
\end{equation}
Truncated exponential distribution is used for sampling distribution $P$, which truncates exponential distribution in $[0, S_M - S_m)$. Note that if uniform distribution is used for $P$, this sampling method will degrade to one of the sampling method proposed by Jarabo et al. \cite{jarabo2014framework}, yet their purpose is to uniformly distribute path vertices. However, since the transmittance for the connection path is $\exp(-\sigma_t S)$, the truncated exponential distribution favors paths with higher overall transmittance, leading to more samples with lower $S$ values. This aligns with our goal of importance sampling based on overall radiance contribution. The sampling PDF takes the form as Equation \eqref{eqn:path-con-pdf}, and full derivation can be found in supplementary note (Section A.2).
 \begin{equation}\label{eqn:path-con-pdf}
 \begin{aligned}
p(t) = p(S)\biggl(\frac{dt}{dS}\biggr)^{-1} =\\ \frac{\eta(S - t)\sigma_t\exp(-\sigma_t (S - S_m))}{c(S - C\cos\theta)(1 - \exp(-\sigma_t (S_M - S_m))},
\end{aligned}
\end{equation}
The other case involves surface events within sampling range, as shown in the right half (case II) Figure \ref{fig:figure6} (b). The detection for this case is performed by comparing the closest surface distance $t_{s}$ with the polar distance $t_{M}$ concerning the ellipse defined by $S_M$. Since sampling beyond surfaces is not feasible, we first sample scattering events as in Section \ref{sec:sec-4-3}, with $p_m = t_{\text{vol}} / (t_{\text{vol}} + t_{\text{surf}})$. For a surface event, we directly output $t_s$, while for a medium event, elliptical sampling is employed and the upper bound for $S$ is updated according to $t_s$. 
\begin{figure}[htp]
\centering
\subfloat[histogram comparison]{\includegraphics[width=0.58\linewidth]{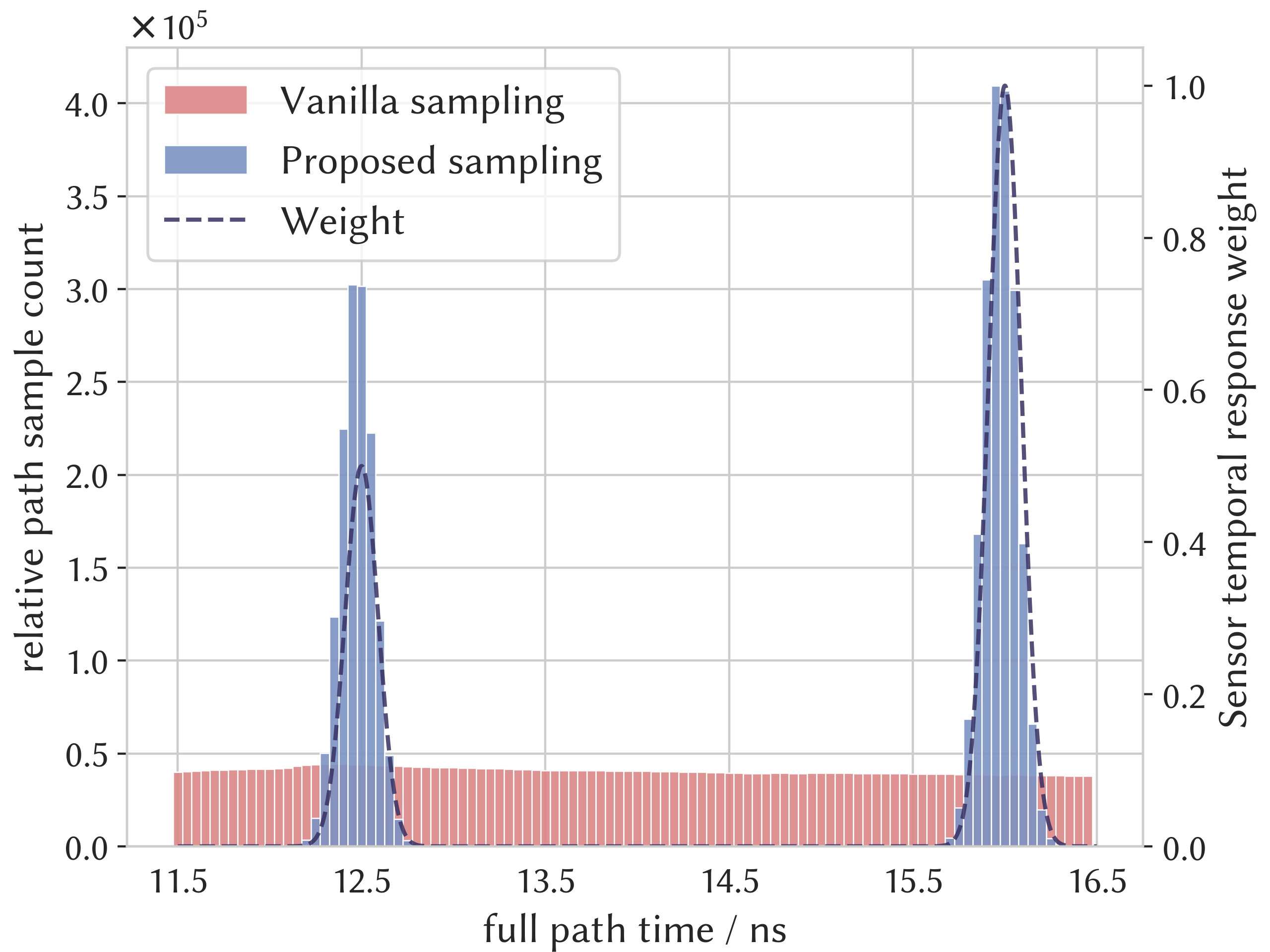}}
\subfloat[rendering comparison]{\includegraphics[width=0.41\linewidth]{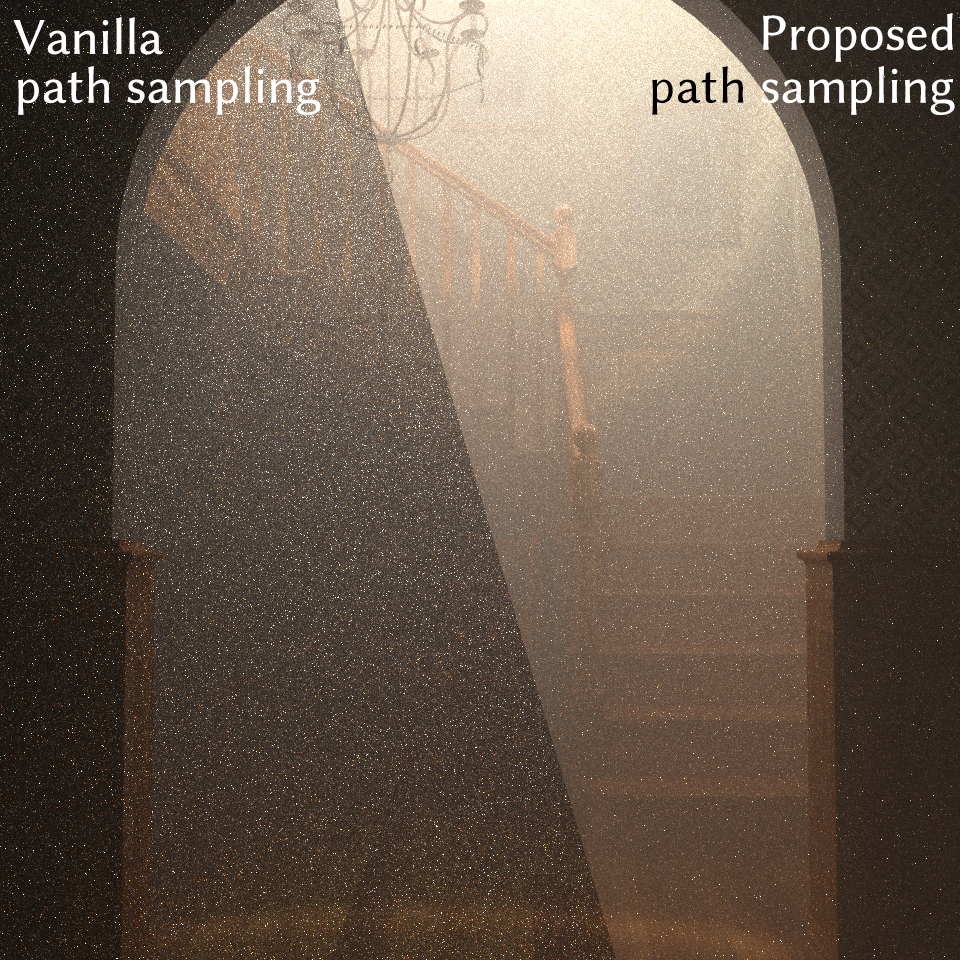}}
\caption{Importance sampling temporal response function. We record full path samples in a temporal histogram during rendering. Histogram (a) shows that the \textcolor[HTML]{DC8686}{vanilla sampling} method struggles to generate path samples in time intervals with higher sensor response weights, in contrast to the \textcolor[HTML]{6B82B2}{proposed sampling method}. Consequently, the vanilla sampling approach generates a substantial number of path samples with little overall contribution, resulting in considerably noisier rendered output (b, left half). The temporal response weight comprises two peaks, and two brighter rings are noticeable in the rendered image (b), one on the floor and the other on the wall. }
\label{fig:figure7}
\end{figure}

We summarize the abilities of our elliptical sampling method: first, it is able to get the upper bound for distance samples for any residual time via Equation $\eqref{eqn:ell-samp}$, which helps to identify non-causal samples mentioned in Section \ref{sec:sec-4-3}; second, the upper bound can also be used for the early identification of paths exceeding the target time range, leading to earlier exit for the path construction and further accelerate rendering by around 1.5 times. Also, its path length control ability can prevent sample rejection: for any given path $\overline{\mathbf{x}}$, the path sampling PDF $p(\overline{\mathbf{x}})$ is non-zero if and only if the temporal response weight $W(\Vert \overline{\mathbf{x}} \Vert)$ is non-zero with the visibility term ignored. This ensures optimal path length control and can entirely avoid sample rejection caused by $W(\Vert \overline{\mathbf{x}} \Vert)$, given the time interval to be sampled from. We present an example of importance-sampling a known $W(\Vert \overline{\mathbf{x}} \Vert)$ with two peaks. The results presented in Figure \ref{fig:figure7}  verify the effectiveness of the proposed method: the two peaks of $W(\Vert \overline{\mathbf{x}}\Vert)$ are depicted with dashed line in Figure \ref{fig:figure7}(a). The vanilla method refers to the direct ToF extension of steady state method. Since the vanilla sampling method relies on direct shadow connection, it fails to generate samples falling within the regions with high weights, resulting in a significant portion of samples (85.31\%) being rejected due to zero weight. In contrast, our method is able to capture the shape of the response weight and thus avoids sample rejection (<3\%). The rendering output in Figure \ref{fig:figure7}(b), which contains two bright rings, shows that our method significantly improves rendering quality. We also provide a proof regarding the optimality of the proposed path control in our supplementary note (Section A.3).

\section{Results and Evaluation}

\begin{figure*}[htp]
    \centering
    \subfloat{
        \centering
        \includegraphics[width=0.99\textwidth]{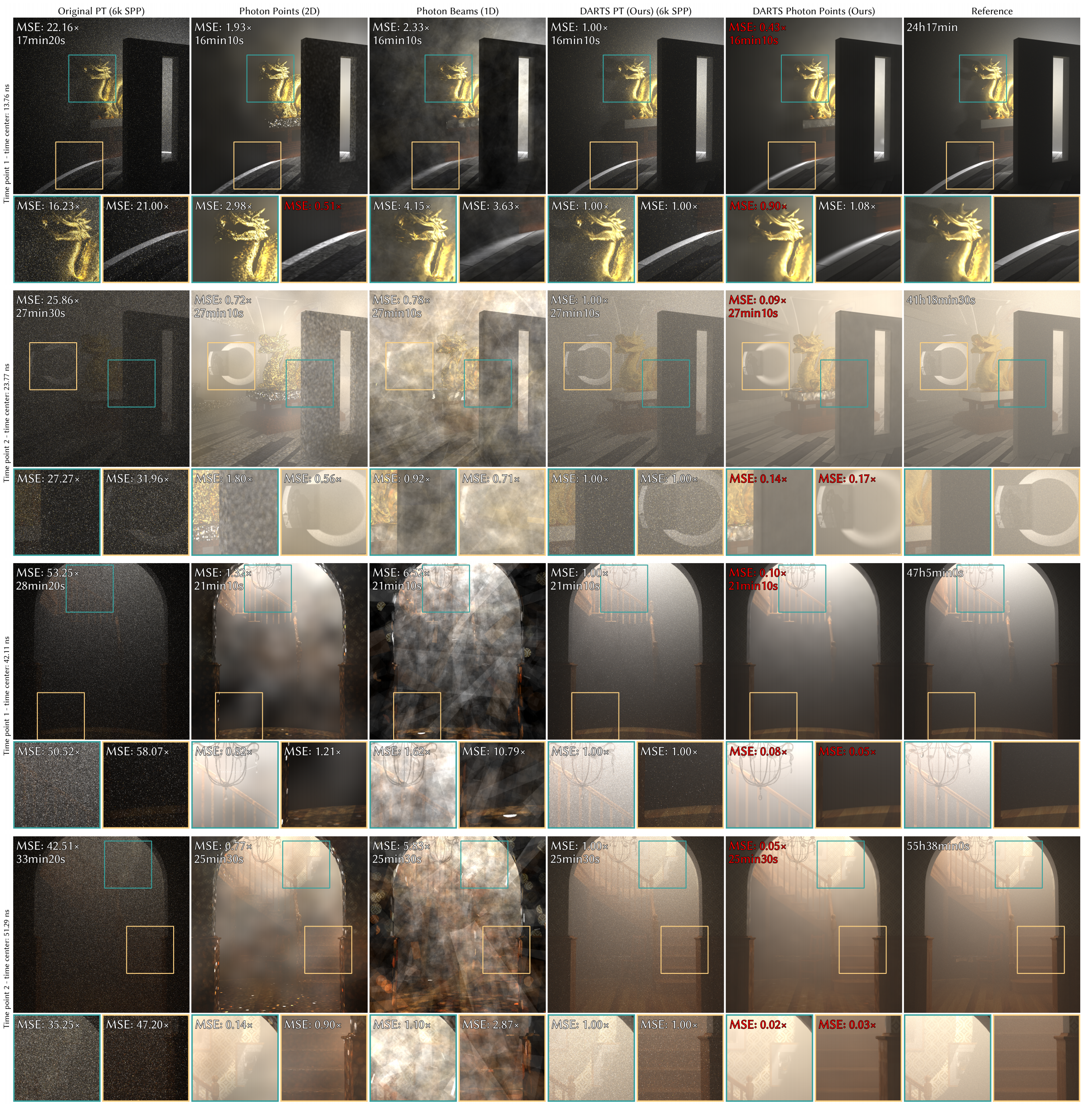}
    }
    
\caption{Time-gated rendering experiments. For each scene, we present the time-gated images rendered by five different methods (organized by column) in two different time intervals (shorter and longer target path lengths). Each image will have two selected areas displayed in magnified views. For original and the proposed methods, we use 6k SPP to render each image and the photon based methods have rendering time equal to the proposed method. We set the MSE of our DARTS PT to be 1, and MSE relative ratios and rendering time are displayed on the top right of each image. The best statistics (row-wise) are \textcolor[HTML]{FE0000}{highlighted in red}. The leftmost texts describe the time point and the corresponding interval center of the rendered images. The original outputs are in HDR format, we therefore normalize the images with their 0.99 quantiles and clamp the output to [0, 1]. It can be seen that our DARTS PT (4th column) and DARTS photon points (5th column) methods have greatly improved rendering quality, compared to their original counterparts. Notably, photon points equipped with DARTS can achieve noise-free rendering with no obvious visual artifacts.}
\label{fig:disp-time-gate-exp}
\end{figure*}

\begin{figure*}[htp]
    \centering
    \subfloat[pixel patch position]{
        \includegraphics[width=0.18\textwidth]{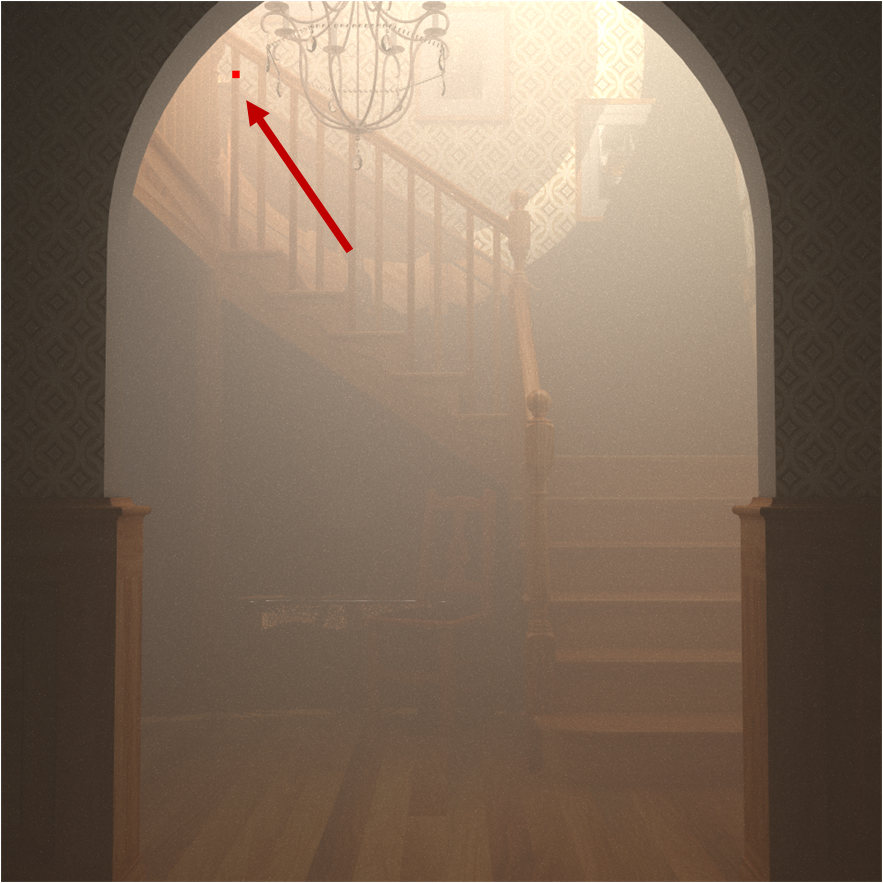}
    }
    \subfloat[PT method comparison (same 80k SPP)]{
        \includegraphics[width=0.4\textwidth]{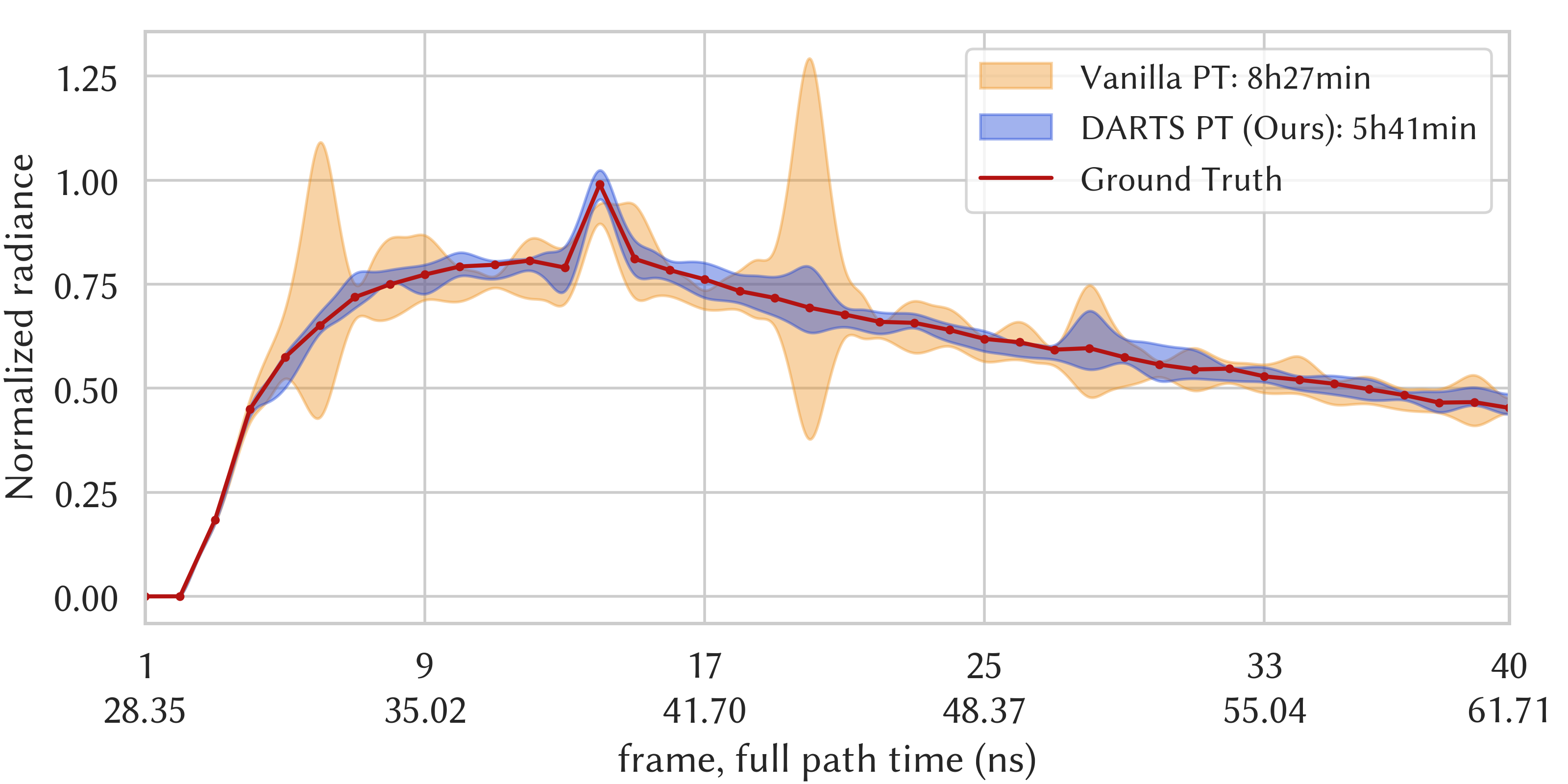}
    }
    \subfloat[PP method comparison (same time, 1h)]{
        \includegraphics[width=0.4\textwidth]{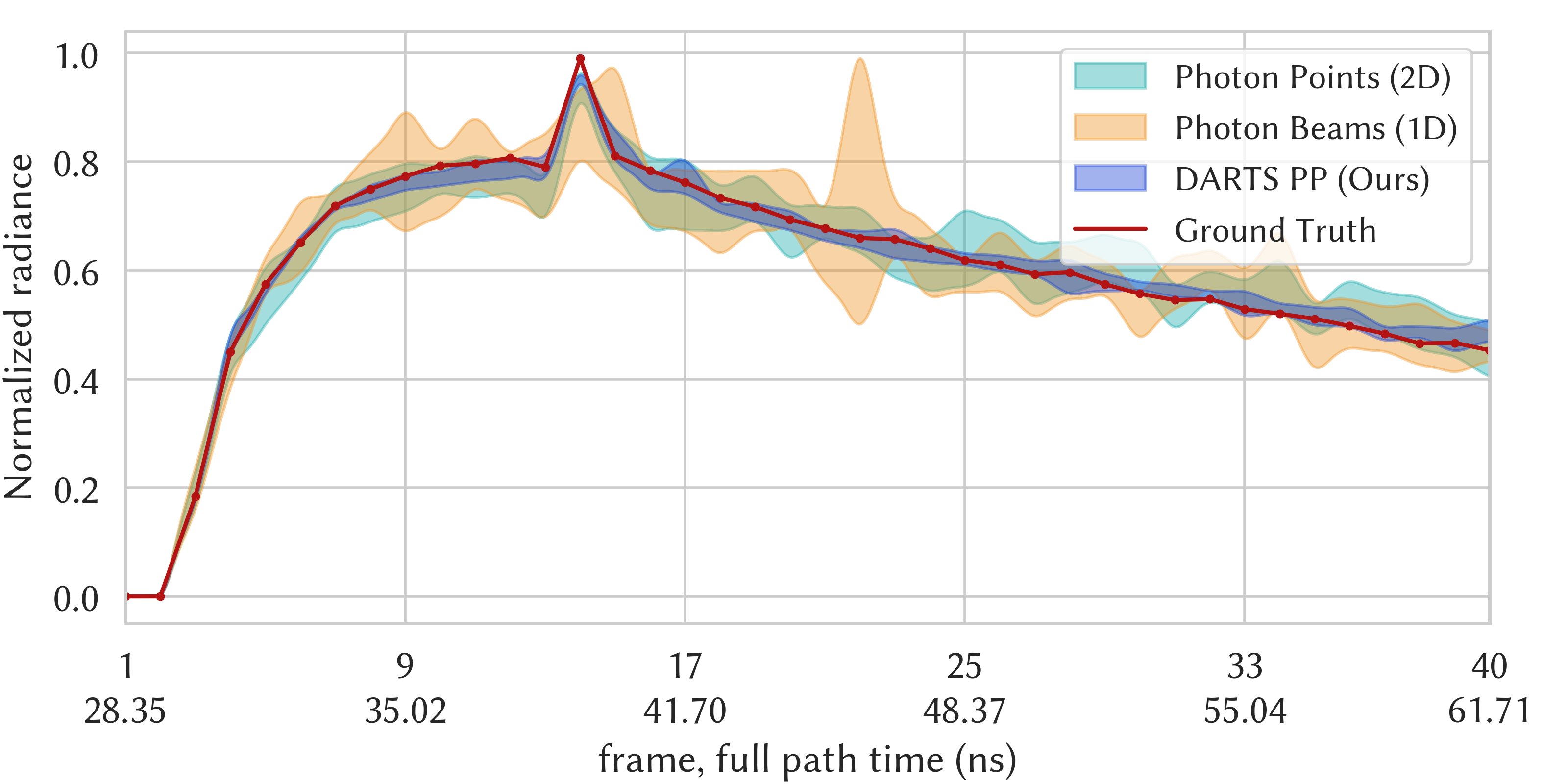}
    }
\caption{Transient rendering experiments. We present the transient curve produced by taking the average of the $6\times 6$ pixel patch in (a). The error distribution curves are smoothed with spline interpolation. Equal-SPP comparison between the vanilla renderer and our DARTS based one is given in (b), with the rendering time given in the graph legend. Equal-time (1h) comparison between vanilla photon based methods and our DARTS based photon point is given in (c). Both curves show that DARTS is able to reduce the asymptotic variance and improve the rendering quality. }
\label{fig:trans-rendering-exp}
\end{figure*}

We compare the proposed method with the vanilla path tracing (PT) and the photon based methods \cite{marco2019progressive, liu2022temporally}, which are the state-of-the-art methods in ToF rendering. All the photon based methods being compared are progressive. We present time-gated rendering and transient rendering experiments in Section 6.1 and 6.2, respectively. Additionally, we provide numerical results in Section 6.3 to demonstrate the stability of our algorithm under various scene and sensor settings. We implement our sampling method in the unidirectional PT of \textit{pbrt-v3} \cite{pharr2023physically}, photon points and photon beams (PP and PB for short, respectively) methods of \textit{Tungsten} \cite{Bitterli:2018:Tungsten}. These renderers are verified to converge to the same results with correct settings and code adjustments (Section B.1 of our supplementary note). Unlike other works that mostly focus on volumetric transport and low order scattering, our experiment setup enables full transport and great maximum allowed depth (such as 200 bounces). All the experiments are done on 112-core Intel Xeon Platinum 8280 CPU@2.60GHz with 104 threads. GPU used for tabulation is Titan RTX. We also run the rendering to converged state to ensure that DARTS remains unbiased and consistent. All rendered HDR images are in the linear RGB color space unless stated otherwise. The reference images are produced by path tracing method.

\begin{table}
    
  \caption{Properties of the test scenes. Note that scattering intensity is measured by the camera-to-emitter transport mean-free-path (TMFP) distance. Higher value means more expected scattering events along the distance.}
  \footnotesize
  \label{tab:scene-prop}
  \begin{tabular}{lll}
    \toprule
    Properties \textbackslash\;Scenes&GLOSSY DRAGON& STAIRCASE \cite{bitterli2016}\\
    \midrule
    Triangle Count& 87k& 262k\\
    PT rendering SPP & 6k& 6k\\
    Scattering intensity* & 3.2 TMFP& 4.3 TMFP\\
    Surface material & glossy / specular reflection & plastic coated material\\
  \bottomrule
\end{tabular}
\end{table}

\subsection{Time-gated Rendering}
\label{section:time-gated-exp}

In Figure \ref{fig:disp-time-gate-exp}, we compare our DARTS based renderer with: (1) the vanilla PT (2) 1D progressive transient photon beam method \cite{marco2019progressive} reproduced by Liu et al.\cite{liu2022temporally} (3) Progressive photon point (point-beam 2D, \cite{kvrivanek2014unifying}) implemented by Liu et al \cite{liu2022temporally}. The proposed path sampling method is tested both on PT (denoted by DARTS PT) and photon based method (DARTS photon points, PP for short). For each scene, two time intervals (shorter and longer target path lengths) are rendered and presented as the first and the second row of the related scene, respectively. Note that the estimators proposed by Liu et al. \cite{liu2022temporally} produce results inferior to those listed above in camera-warped and full transport settings, and some of them needs mathematical corrections to work under camera-warped settings, which is out of this paper's scope and are not included (see Figure X in the supplementary note). The presented images are normalized with their 0.99 quantile numbers. In the following, we mainly provide two scenes for comparison. More scenes and results can be found in our supplementary note, interactive local web-viewer and video.

Some of the important properties of the two test scenes: GLOSSY DRAGON (first two rows of Figure \ref{fig:disp-time-gate-exp}) and STAIRCASE \cite{resources16} (last two rows of Figure \ref{fig:disp-time-gate-exp}) are listed in Table \ref{tab:scene-prop}. Both volumetric scenes feature complex geometries and surface transport properties. The pulsed point emitters in each scene are not within the direct line-of-sight. We render the scene with vanilla approach and DARTS with equal sample counts (6k SPP), and the photon based methods employ the same rendering time as DARTS PT. Quantitative results presented in Figure \ref{fig:disp-time-gate-exp} are normalized by the MSE of DARTS PT results.

It is evident that our DARTS PT significantly outperforms the vanilla PT in output quality and even requires less time to render with the same SPP, reducing MSE metrics by about 10-50 times. Moreover, DARTS PT can already outperform photon based methods in most scenes due to its bias-free nature, while DARTS PP further increased the gap between our methods and the compared methods. We also observe that, though progressive method is employed, rendering under a fixed time budget and accounting for full transport introduces bias that proves challenging to mitigate, even with extensive parameter tuning. This challenge is particularly pronounced in the case of the photon beams method, leading to noticeable visual artifacts and a performance drop. With our sampling method, photon based methods can adopt lower photon counts per sample to achieve higher SPP and the parameters are easier to tune, and thus have better convergence.

\subsection{Transient Rendering}
Our method is better suited for time-gated rendering where temporal path reuse \cite{jarabo2014framework} isn't employed, but we still report performance improvements in transient rendering using our proposed methods. Similar to time-gated rendering, we compare PT and PP with our sampling method against their vanilla counterparts and vanilla photon beams. Here we present comparison experiment for STAIRCASE scene. Each frame has the same temporal width as time-gated experiments and 40 frames instead of only 1 frame are rendered. Since it is generally challenging for PT to produce transient images with low variance, here we compare path tracing methods (rendered with 80k SPP) and photon based method (rendered for the same time, 1h) differently. The transient curve from a $6\times 6$ pixel patch shown in Figure \ref{fig:trans-rendering-exp}(a) is presented. The error distribution of the PT method is presented in Figure \ref{fig:trans-rendering-exp}(b), whereas the error distribution of the photon-based method is depicted in Figure \ref{fig:trans-rendering-exp}(c).

The results presented in Figure \ref{fig:trans-rendering-exp} show that compared to vanilla methods, the proposed method is able to improve the rendering quality with significantly less rendering time. However, photon based methods outperform path tracing based renderer with the proposed sampling methods, due to the utilization of temporal path reuse and spatial-temporal blurring (rectangle kernel for temporal blurring). We further test the photon points methods with our proposed sampling methods, observing improvements in both rendering quality and rendering speed. 

The reduction in rendering time is achieved by eliminating temporal path reuse: the proposed method constructs paths and connections for a specific frame, and the subsequent random walk are terminated once the path time exceeds the duration of that specified frame, thus requiring fewer overall bounces. The rendering quality is improved due to the ability of importance-sampling the whole path according to transient radiance contribution. Besides, the proposed methods are able to uniformly distribute the path samples in time-domain, similar to the work of Jarabo et al's \cite{jarabo2014framework} but in a different way: our work is able to place equal number of path samples in each frame (statistically, regardless of visibility term) and our constructed path is not reused across frames but dedicated to a single frame instead. Thus, our path samples don't need to compromise among all the frames.  

\subsection{Parametric Variation and Further Comparisons}
In the following, we mainly discuss the performance of our method under different scattering coefficient $\sigma_s$, temporal gate width and total rendering time. This section concludes with an ablation study of our sampling method in the CORNELL BOX scene. Unless otherwise specified, all subsequent experiments employ 2k SPP for PT, while photon-based methods adhere to the same rendering time as DARTS PT. Rendered figures are available in the supplementary note and our interactive local web-viewer.

\subsubsection{Scattering coefficient} 
\begin{figure}[htp]
    \centering
\includegraphics[width=0.99\linewidth]{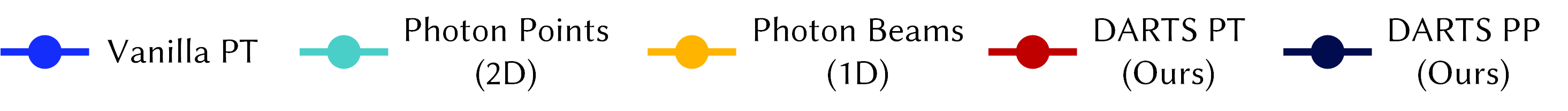}\vspace{-1em}
    \subfloat[GLOSSY DRAGON time point 1]{
        \centering
        \includegraphics[width=0.49\linewidth]{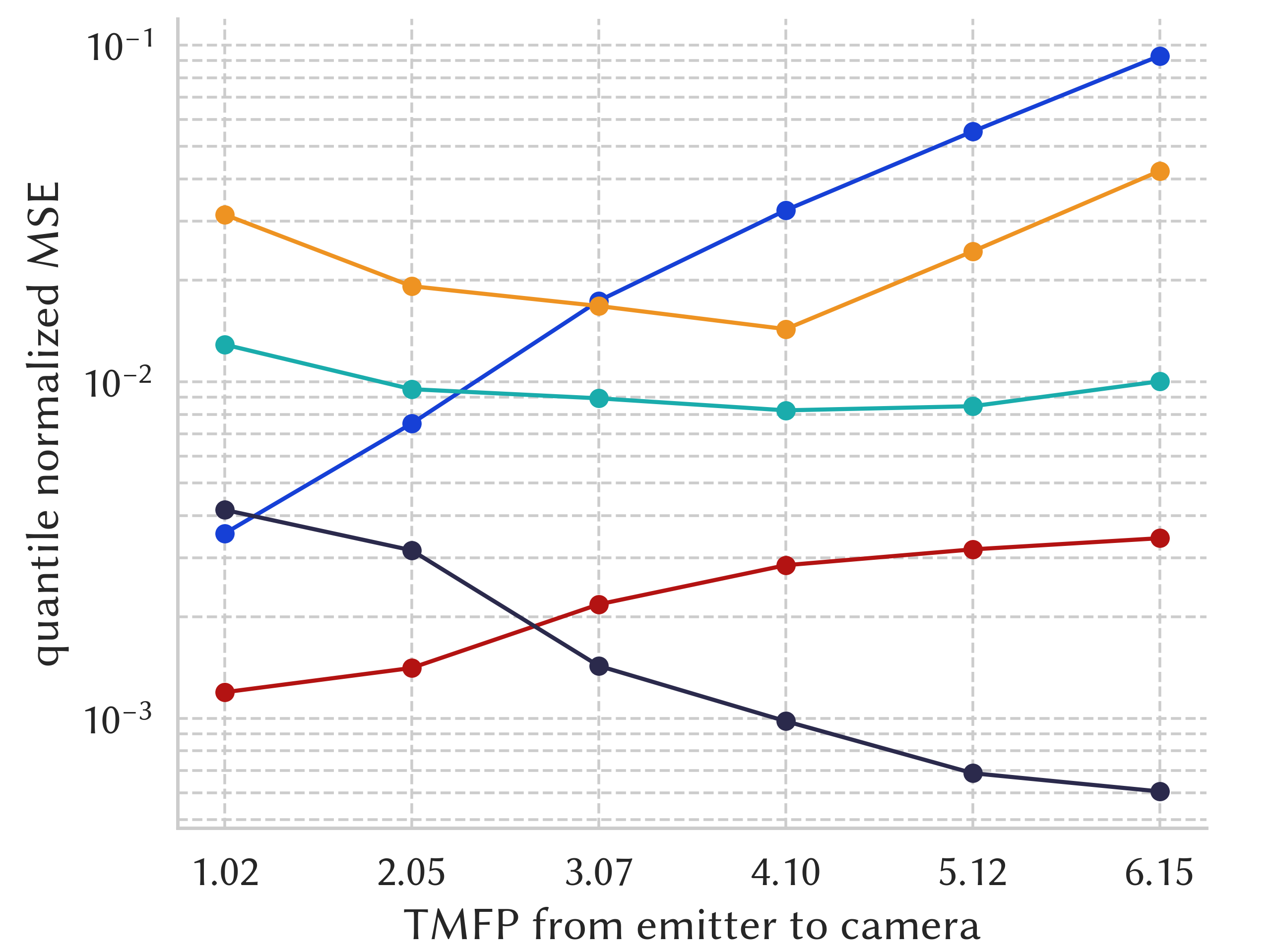}
    }
    \subfloat[GLOSSY DRAGON time point 2]{
        \centering
        \includegraphics[width=0.49\linewidth]{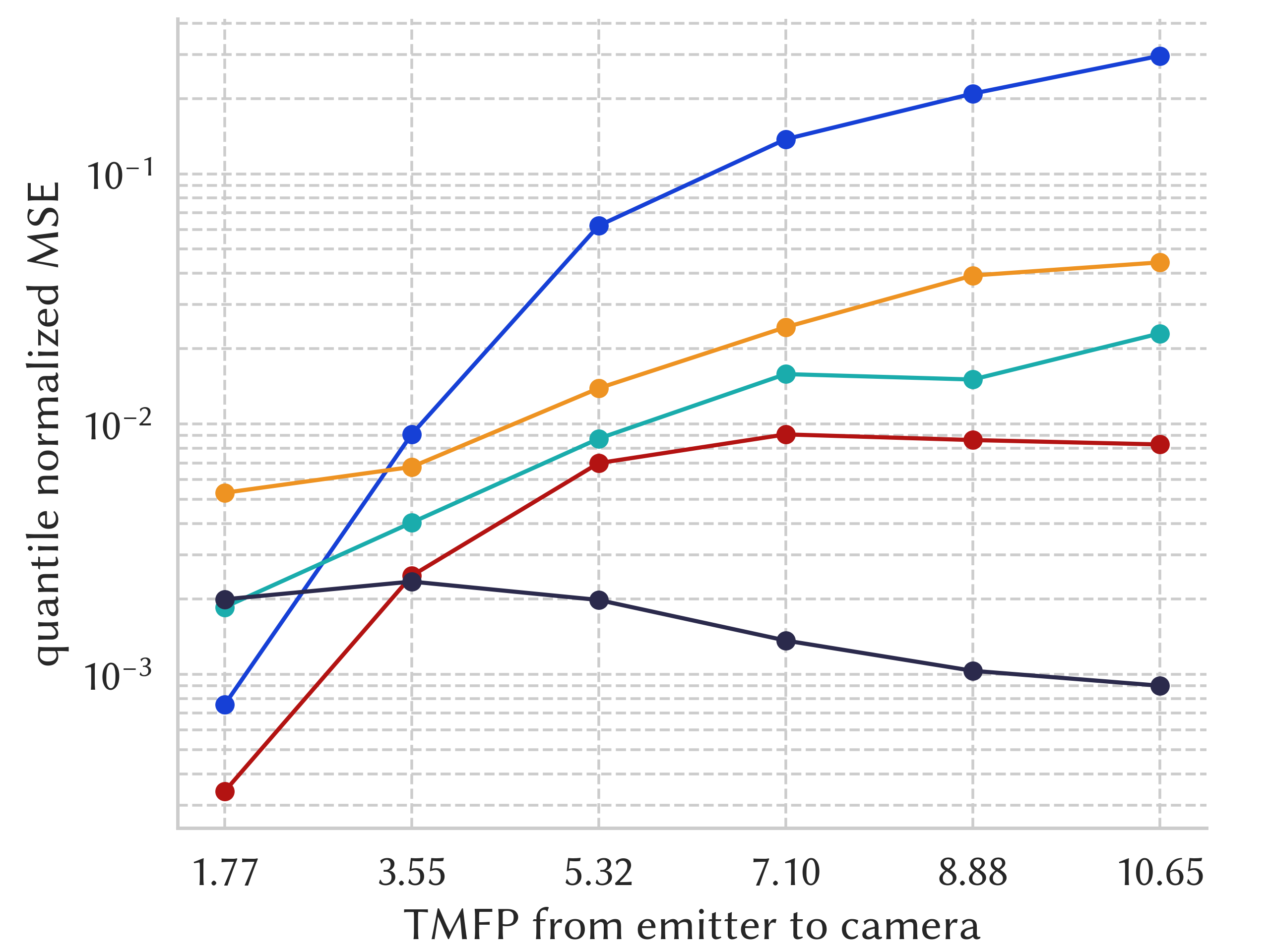}
    }\\[-2ex]
    \subfloat[STAIRCASE time point 1]{
        \centering
        \includegraphics[width=0.49\linewidth]{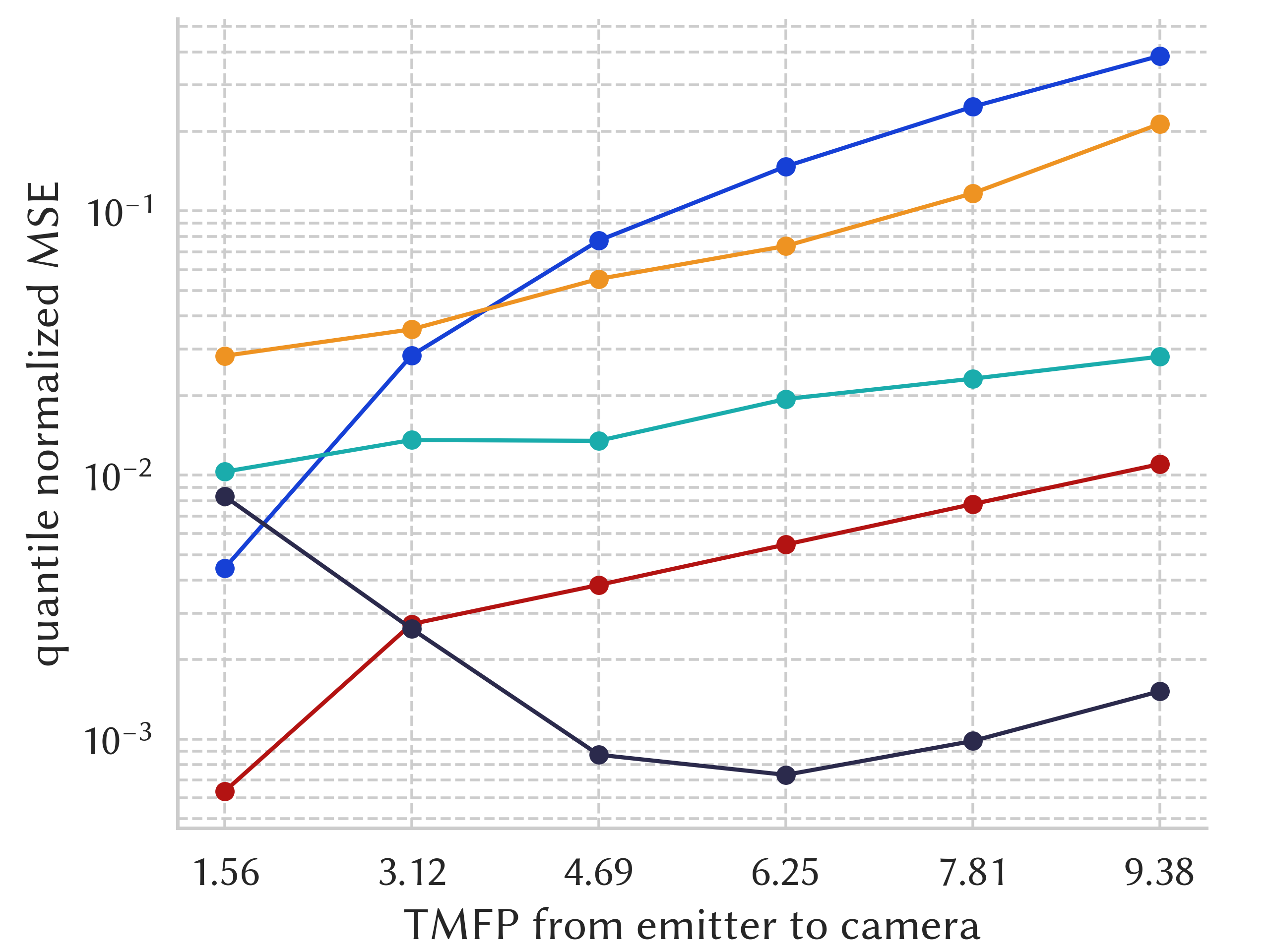}
    }
    \subfloat[STAIRCASE time point 2]{
        \centering
        \includegraphics[width=0.49\linewidth]{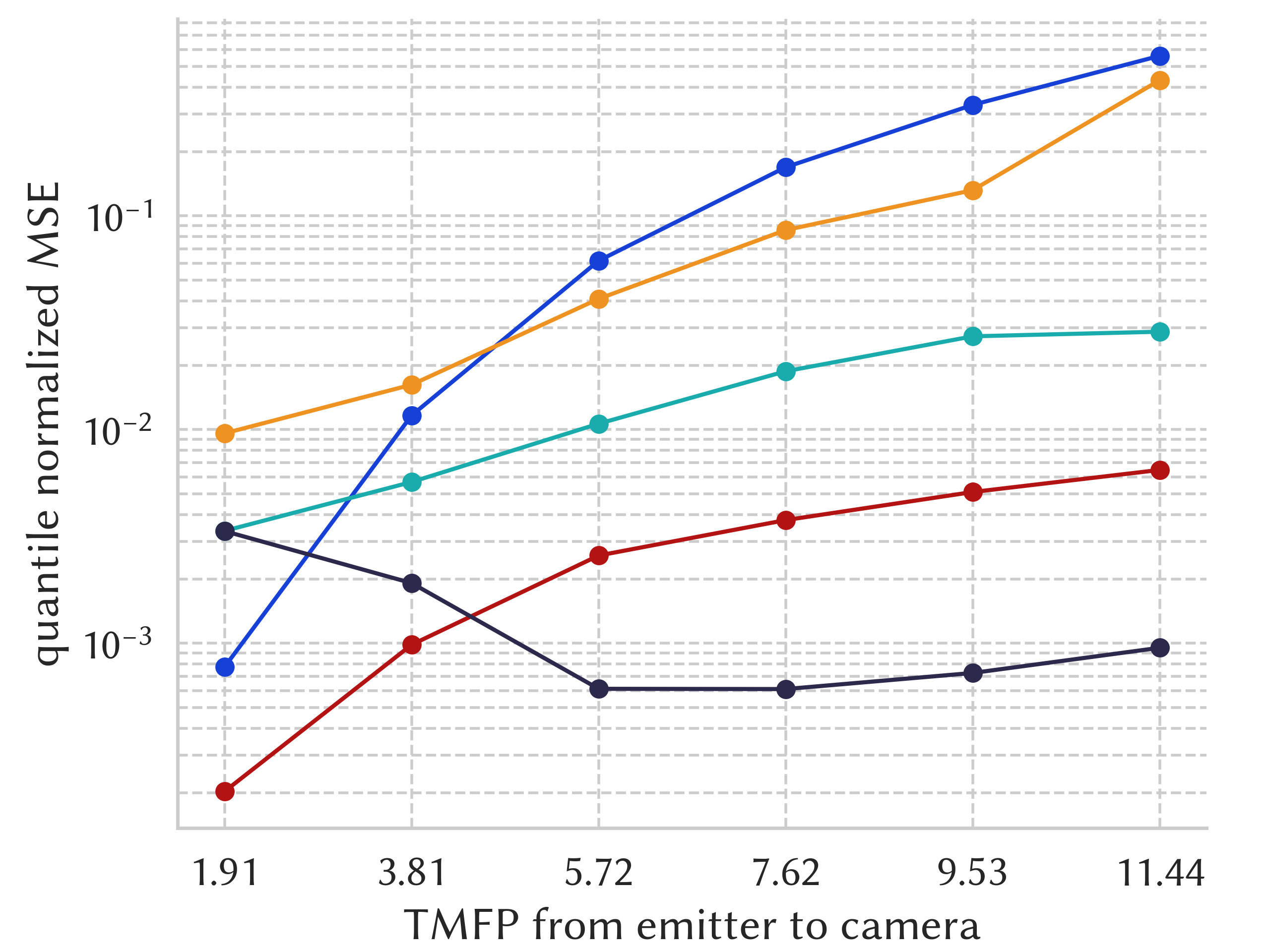}
    }
    \caption{Scattering  coefficient \& MSE relationship curves. Curves depicting the relationship between scattering coefficient (converted to TMFP) and MSE are presented. The first and the second row are obtained from GLOSSY DRAGON and STAIRCASE scene, respectively. Note that multiple scattering at higher $\sigma_s$ values contributes to an overall decrease in signal amplitude and, consequently, a reduction in MSE. To account for this, we normalize the rendered results using the 0.95 quantile before MSE calculation.}
    \label{fig:scat-coeff-curves}
\end{figure}

For both GLOSSY DRAGON and STAIRCASE scene, we employ six different $\sigma_s$ settings, ranging from low-order to high-order scattering. In each scene, we select the same two time intervals as in the time-gated rendering experiments for rendering, and 30 images are rendered to calculate MSE. For the sake of clarity, we convert the scattering coefficient to TMFP values for path length of the corresponding time interval. To obtain reasonable outputs, all photon-based methods require parameter tuning for different scattering coefficients. In the following figures, we present the comparison results in two above-mentioned scenes. It can be seen from Figure \ref{fig:scat-coeff-curves} that our DARTS PT maintains lower variance compared to both vanilla PT and photon-based methods, exhibiting greater stability when the scattering coefficient varies. Meanwhile, DARTS PP gradually gains its superiority as the scattering media get denser but in thin scattering media, its performance is inferior to path tracing methods, due to the surface transport. Compared with the experiments presented in Section \ref{section:time-gated-exp}, photon-based methods employ less rendering time, leading to fewer SPP, which can impact progressive rendering and result in a performance drop. 

\subsubsection{Temporal gate width} 
\begin{figure}[htp]
    \centering
\includegraphics[width=0.99\linewidth]{figures/curve-legend-labels.png}\vspace{-1em}

    \subfloat[t][STAIRCASE time point 1]{
        \centering
        \includegraphics[width=0.49\linewidth]{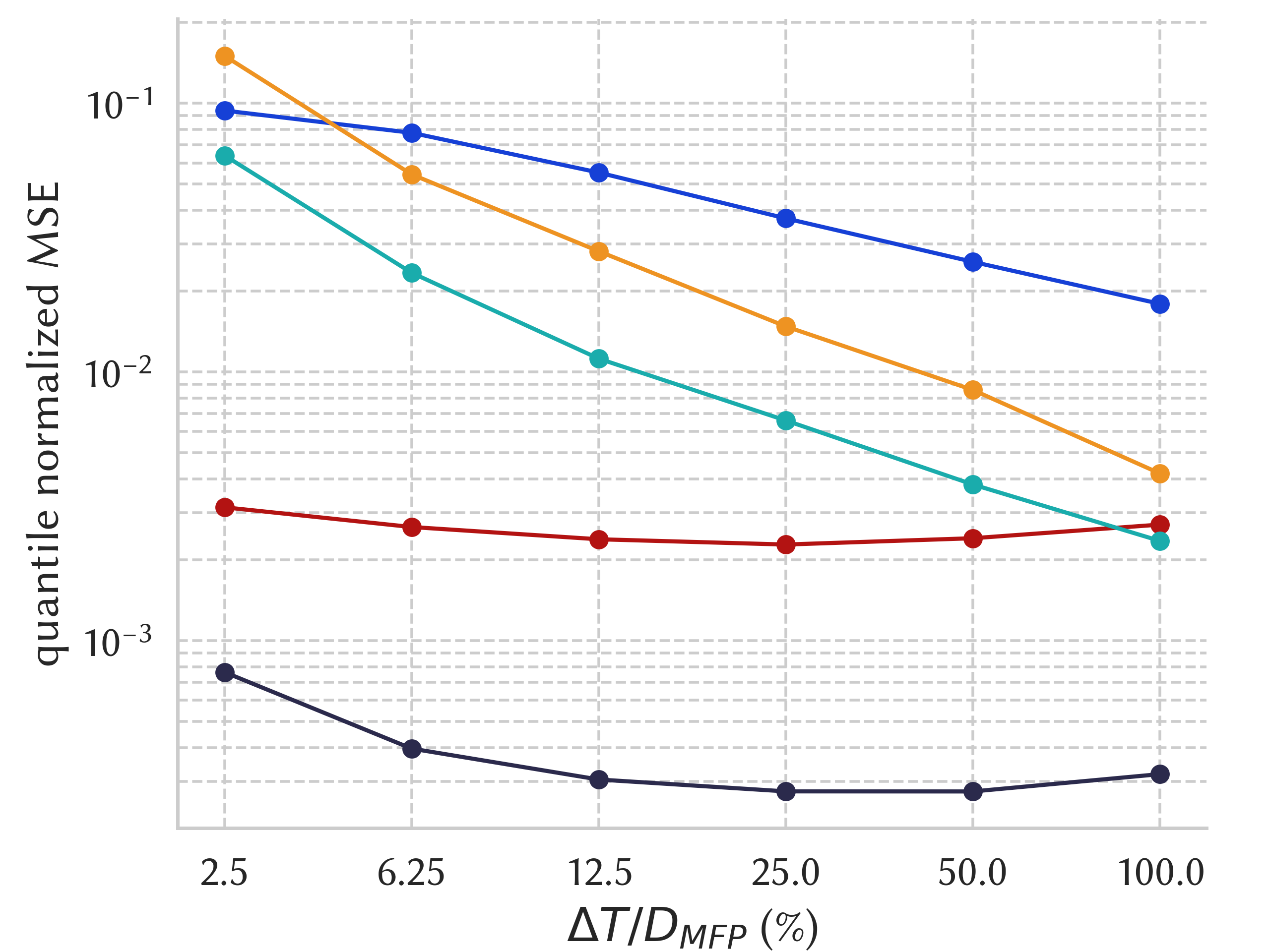}
    }
    \subfloat[t][STAIRCASE time point 2]{
        \centering
        \includegraphics[width=0.49\linewidth]{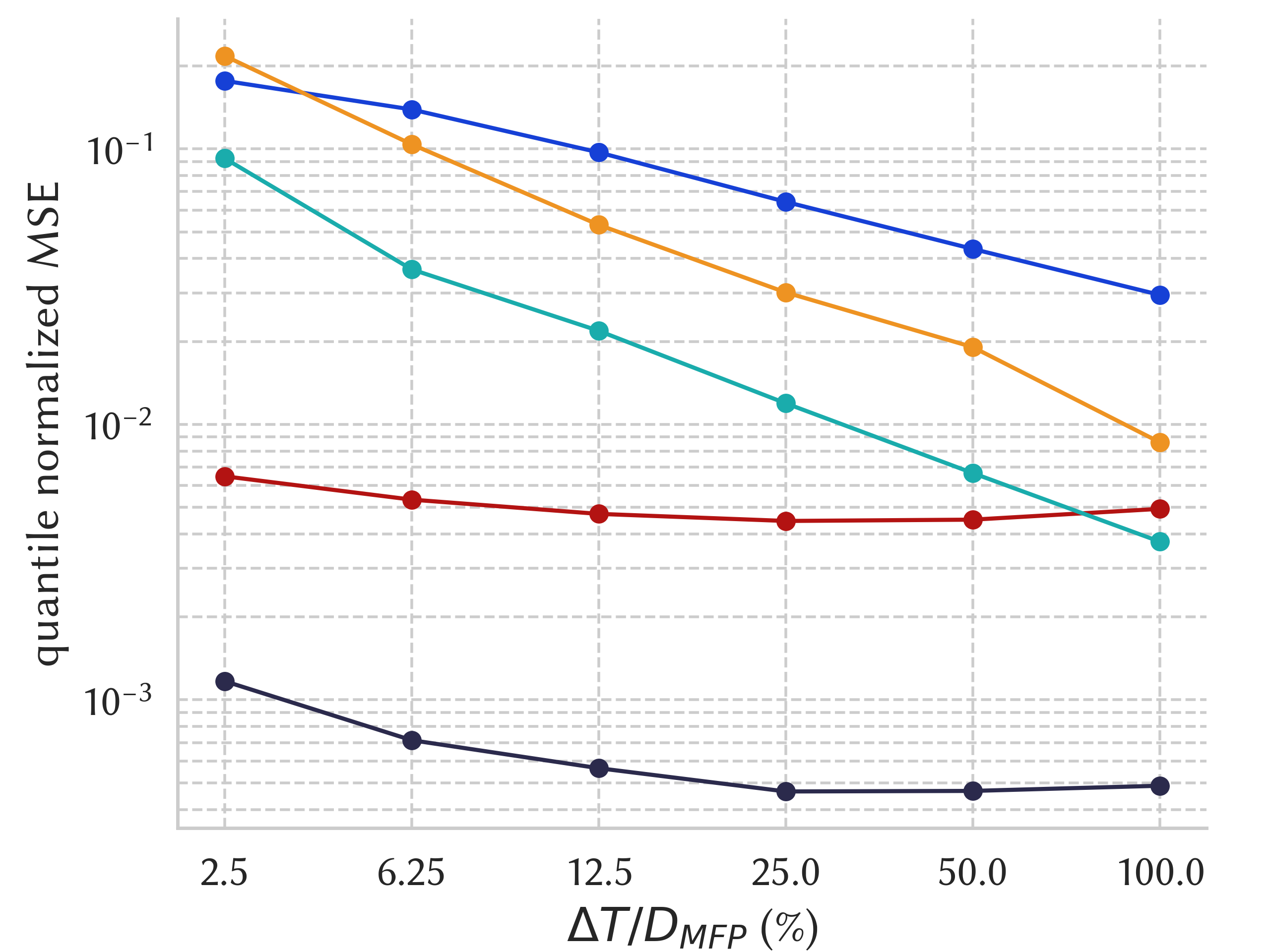}
    }
    \caption{Temporal gate width \& MSE relationship curves. The test scene is STAIRCASE scene and the outputs are also normalized with 0.95 quantile. Our proposed method is seen to be stable across different settings.}
\label{fig:time-gate-width-curve}
\end{figure}

Here we present the gate width experiments in the STAIRCASE scene. Six gate widths are selected, ranging from a narrow delta-function-like gate to a wide steady-state-like gate. Likewise, the gate width $\Delta T$ is converted into a ratio with the mean free path (MFP) of the scene. The scene settings align with those of time-gated experiments described in Section \ref{section:time-gated-exp}. The curves are given in Figure \ref{fig:time-gate-width-curve}. We can observe that for wider time gates, as the setup more closely resembles steady-state rendering, the advantages of DARTS PT in time-resolved sampling are less observable over path tracing methods. Therefore, the gap between the DARTS PT and photon points methods is gradually narrowing. 
\subsubsection{Numerical convergence} 
\begin{figure}[htp]
\includegraphics[width=0.99\linewidth]{figures/curve-legend-labels.png}

\includegraphics[width=0.95\linewidth]{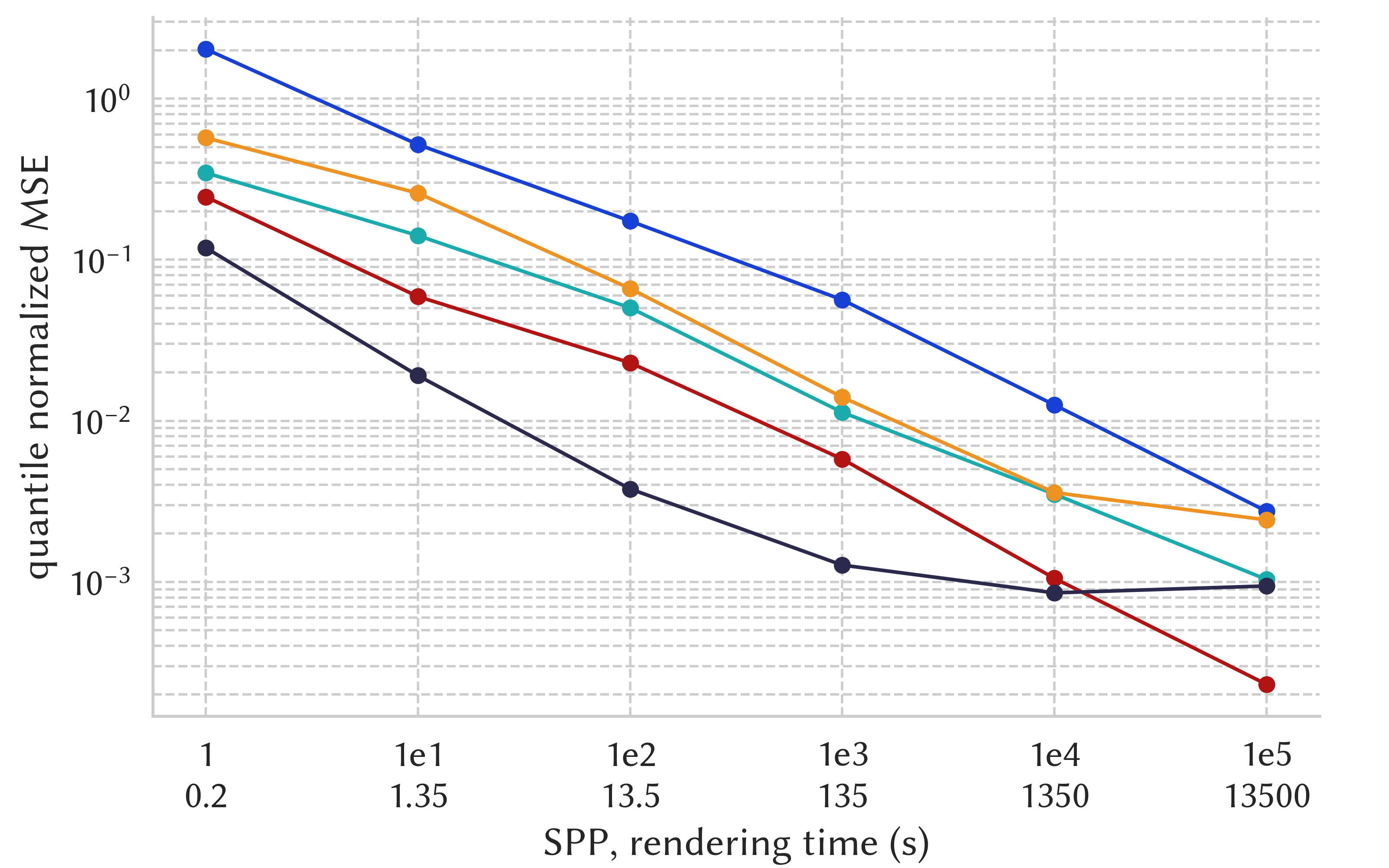}
\caption{Numerical convergence curve. We compare the proposed method with vanilla method and photon based methods under several different SPP (thus, rendering time) settings in GLOSSY DRAGON scene. Both SPP (for path tracing methods) and rendering time (for both path tracing and photon based methods) are presented in the x-axis labels. Note that vanilla PT uses the same number of SPP as DARTS PT, and the vanilla PT takes approximately 1.5 times longer to render compared to DARTS PT.}
\label{fig:converge-curve}
\end{figure}

A straightforward numerical convergence analysis is depicted in Figure \ref{fig:converge-curve}, where different SPP settings (scaled by a factor of ten) are employed to render the GLOSSY DRAGON scene. For photon based methods, we have rendered the scenes with rendering time equivalent to DARTS PT. The result shows that: for PT methods, the proposed path sampling method is able to improve the MSE convergence by at least an order of magnitude in this scene; for photon based method, the convergence for fewer-sample cases are significantly improved, whereas as the SPP increases, the improvement is less significant and the performance can even be inferior to DARTS PT. This is caused by the inherent bias of photon based methods, which is difficult to eliminate, even though the methods are progressive.
\subsubsection{Ablation study} 
\begin{figure*}[h!tb]
\subfloat[Vanilla Method]{\includegraphics[width=0.24\textwidth]{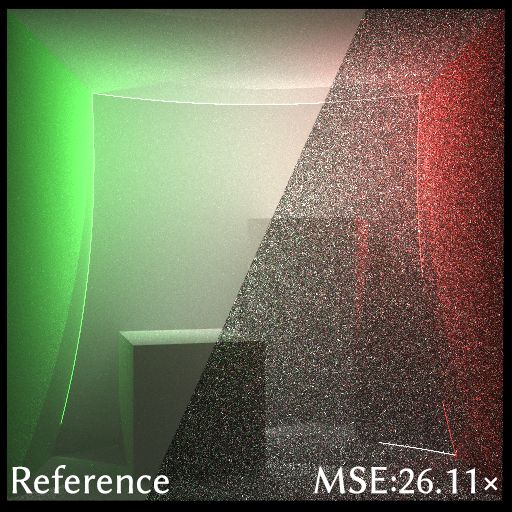}}\;
\subfloat[DA distance sampling only]{\includegraphics[width=0.24\textwidth]{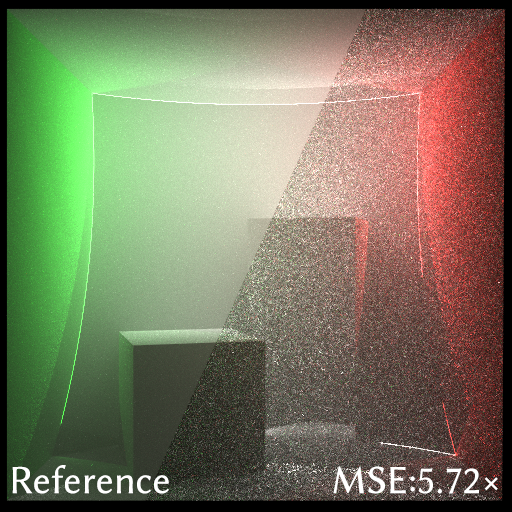}}\;
\subfloat[EDA sampling only]{\includegraphics[width=0.24\textwidth]{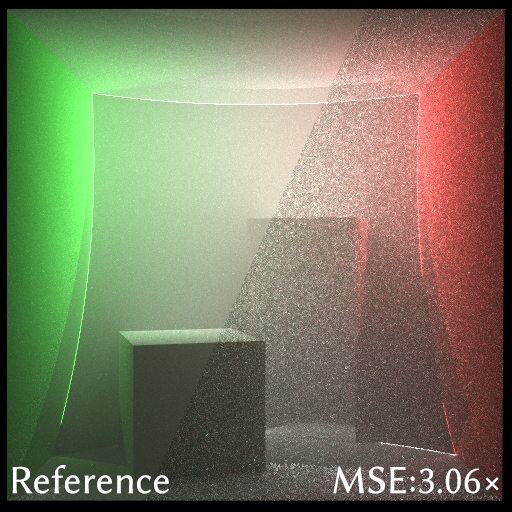}}\;
\subfloat[Full DARTS]{\includegraphics[width=0.24\textwidth]{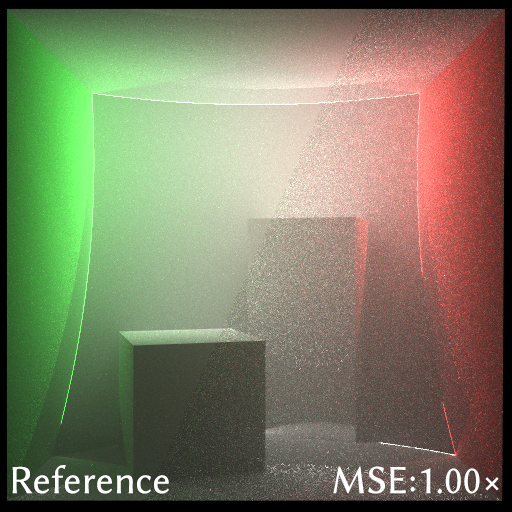}}
\caption{Ablation study: an example case. Comparison between DARTS and three alternative strategies is presented under the same rendering time settings. The images shown above are synthesized by averaging 40 images (totaling 30 min for each case), with the reference being rendered for 4.5h. Images shown are normalized by their 0.99 quantile. We normalize all the MSE results by the MSE of images produced by DARTS. In each image, the left half is the reference and the right half employs strategies as indicated by the sub-captions. The proposed method results in a substantial reduction in MSE (in this case, variance).}
\label{fig:ablation-examples}
\end{figure*}

Our proposed sampling method consists of two parts: DA distance sampling and EDA sampling. We validate the ability of both sampling methods in improving rendering quality through four strategies: vanilla renderer, the renderer with only one of the proposed methods and the renderer with both methods enabled. Here we use a modified version of the CORNELL BOX scene with a pulsed point source. The scene is rendered under various scene settings such as different scattering coefficients and temporal gate widths. Both qualitative and quantitative results under one example setting are presented in Figure \ref{fig:ablation-examples}. It is evident that the collaborative utilization of both proposed methods substantially improves rendering quality. For additional results of other settings, one can refer to our supplementary materials (Figure III in Section C.1).

\section{Discussion and Limitations}
In summary, we have proposed a novel path sampling method to improve the overall quality of ToF rendering tasks. The proposed method is derived based on diffusion approximation in homogeneous scattering media and its Monte Carlo integration in a residual-time-defined ellipse, together with the non-trivial extension of ellipsoidal connection that can be directly applied for scattering media. Our experiments demonstrate that the proposed method is able to improve rendering quality and efficiency in both path tracing and photon-based methods. Moreover, the improved path tracing method performs comparably to, and in some cases surpasses, photon-based methods in scenes with scattering media and complex surface properties. We anticipate that our work will contribute to the simulation of increasingly sophisticated ToF sensors in the field of optics and sensing. 

In the following, we first discuss the implementation differences of the proposed method in different frameworks, and then provide the limitations and the future avenues for this work.

{\fontfamily{LinuxBiolinumT-OsF}\selectfont
\textit{Implementation differences in different frameworks.}
}The previous derivations mainly stem from PT framework. While the same methodology can be applied to photon-based method, we do highlight two major differences in implementation: (1) DA distance sampling and EDA sampling are only used in photon pass, and the camera will be considered as a virtual importon \cite{vorba2014line} emitter, while in PT, since the path starts from the camera, the real emitter is used. (2) In photon based methods, we cache the control vertex in the photon map and leave the gathering for the sensor pass, while in PT, a generalized shadow connection will be made to evaluate the radiance immediately after sampling a control vertex. For detailed explanation of the differences in implementation, one can refer to Section B.2 in our supplementary note. 

\subsection{Limitations \& Future Avenues}

\subsubsection{Emitter \& phase function types} The emitter in our theoretical derivation is assumed to be a point emitter. We employ two types of emitters in the experiments: point emitters and spot emitters. The latter is employed in photon-based methods, where the camera acts as a spot emitter of importons. For more complex emitters, for example, collimated emitters can be handled by sampling single scattering events and treating each scattering vertex as a virtual point emitter, while the area emitters can be approached by point sampling, with each sample considered as an equivalent spot emitter. The phase function employed is currently assumed to be Henyey-Greenstein phase function, yet we note that our method might be extended to phase functions that have diffusion approximation, such as microflake phase function \cite{jakob2010radiative}. This part is left for future work.

\subsubsection{Heterogeneity \& strongly directed phase function} Our current work addresses the ToF rendering challenges in homogeneous scattering media where the scattering exhibits limited directionality. Media with heterogeneity or strong directionality may violate the assumptions of diffusion approximation. In such cases, diffusion theory may not be applicable when incorporating global transient radiance information. Additionally, as elliptical sampling in EDA involves a two-step approach, the reused direction from EDA may be sub-optimal for peaky phase function theoretically. However, we find that the 2D radiance contribution of the product of two consecutive phase functions for elliptical connection does not differ much from the single phase function case (see supplementary note Section C.5), therefore, local phase function still works well for MIS. For scenes with homogeneous scattering media in distinct regions, our work can still be applied in bidirectional methods, where vertices on emitter paths can be regarded as point sources.

\subsubsection{Scenes with complex visibility} Using diffusion theory in scenes with complex visibility lacks robust theoretical substantiation. In our sampling method, visibility (between emitter and the current vertex) may be left unaccounted for. Our experiments indicate that, even when emitters are not directly visible, our sampling method can still greatly enhance rendering quality. However, incorporating visibility term in unidirectional renderers poses inherent challenges. For example, calculating ray-scene intersection for every candidate sample during RIS becomes computationally burdensome. This overhead might outweigh the performance gains, even if strategies are utilized for acceleration, like caching a shadow map for the emitter.

For future works, it would be interesting to investigate how our method can be applied for differentiable rendering in transient state, since higher sampling efficiency generally yields better backward performance. Additionally, investigating the compatibility of our method with more sophisticated rendering frameworks, such as bidirectional path tracing and metropolis light transport, could prove valuable, particularly in rendering challenging scenes like non-line-of-sight simulation setups. We believe that our study provides inspiration for future researches in time-of-flight sensor simulation, especially in scenarios involving diverse and complex scattering. The testing scenes and code of our method in both \textit{pbrt-v3} and \textit{Tungsten} frameworks can be found in our supplementary material.





\ifCLASSOPTIONcaptionsoff
  \newpage
\fi



%

\bibliographystyle{plain}
\bibliography{refs}

\begin{thebibliography}{10}

\bibitem{ament2014refractive}
Marco Ament, Christoph Bergmann, and Daniel Weiskopf.
\newblock Refractive radiative transfer equation.
\newblock {\em ACM Transactions on Graphics (TOG)}, 33(2):1--22, 2014.

\bibitem{attal2021torf}
Benjamin Attal, Eliot Laidlaw, Aaron Gokaslan, Changil Kim, Christian Richardt, James Tompkin, and Matthew O'Toole.
\newblock T{\"o}rf: Time-of-flight radiance fields for dynamic scene view synthesis.
\newblock {\em Advances in neural information processing systems}, 34:26289--26301, 2021.

\bibitem{resources16}
Benedikt Bitterli.
\newblock Rendering resources, 2016.
\newblock https://benedikt-bitterli.me/resources/.

\bibitem{bitterli2016}
Benedikt Bitterli.
\newblock Virtual femto photography.
\newblock \url{https://benedikt-bitterli.me/femto.html}, 2016.

\bibitem{Bitterli:2018:Tungsten}
Benedikt Bitterli.
\newblock Tungsten renderer, 2018.

\bibitem{bitterli2017beyond}
Benedikt Bitterli and Wojciech Jarosz.
\newblock Beyond points and beams: Higher-dimensional photon samples for volumetric light transport.
\newblock {\em ACM Transactions on Graphics (TOG)}, 36(4):1--12, 2017.

\bibitem{bitterli2020spatiotemporal}
Benedikt Bitterli, Chris Wyman, Matt Pharr, Peter Shirley, Aaron Lefohn, and Wojciech Jarosz.
\newblock Spatiotemporal reservoir resampling for real-time ray tracing with dynamic direct lighting.
\newblock {\em ACM Transactions on Graphics (TOG)}, 39(4):148--1, 2020.

\bibitem{contini1997photon}
Daniele Contini, Fabrizio Martelli, and Giovanni Zaccanti.
\newblock Photon migration through a turbid slab described by a model based on diffusion approximation. i. theory.
\newblock {\em Applied optics}, 36(19):4587--4599, 1997.

\bibitem{deng2019photon}
Xi~Deng, Shaojie Jiao, Benedikt Bitterli, and Wojciech Jarosz.
\newblock Photon surfaces for robust, unbiased volumetric density estimation.
\newblock {\em ACM Transactions on Graphics}, 38(4), 2019.

\bibitem{du2022boundary}
Dongyu Du, Xin Jin, Rujia Deng, Jinshi Kang, Hongkun Cao, Yihui Fan, Zhiheng Li, Haoqian Wang, Xiangyang Ji, and Jingyan Song.
\newblock A boundary migration model for imaging within volumetric scattering media.
\newblock {\em Nature Communications}, 13(1):3234, 2022.

\bibitem{faccio2020non}
Daniele Faccio, Andreas Velten, and Gordon Wetzstein.
\newblock Non-line-of-sight imaging.
\newblock {\em Nature Reviews Physics}, 2(6):318--327, 2020.

\bibitem{georgiev2013joint}
Iliyan Georgiev, Jaroslav Krivanek, Toshiya Hachisuka, Derek Nowrouzezahrai, and Wojciech Jarosz.
\newblock Joint importance sampling of low-order volumetric scattering.
\newblock {\em ACM Trans. Graph.}, 32(6):164--1, 2013.

\bibitem{gruber2019gated2depth}
Tobias Gruber, Frank Julca-Aguilar, Mario Bijelic, and Felix Heide.
\newblock Gated2depth: Real-time dense lidar from gated images.
\newblock In {\em Proceedings of the IEEE/CVF International Conference on Computer Vision}, pages 1506--1516, 2019.

\bibitem{hachisuka2009stochastic}
Toshiya Hachisuka and Henrik~Wann Jensen.
\newblock Stochastic progressive photon mapping.
\newblock In {\em ACM SIGGRAPH Asia 2009 papers}, pages 1--8. 2009.

\bibitem{hachisuka2008progressive}
Toshiya Hachisuka, Shinji Ogaki, and Henrik~Wann Jensen.
\newblock Progressive photon mapping.
\newblock In {\em ACM SIGGRAPH Asia 2008 papers}, pages 1--8. 2008.

\bibitem{9534479}
Abderrahim Halimi, Aurora Maccarone, Robert~A. Lamb, Gerald~S. Buller, and Stephen McLaughlin.
\newblock Robust and guided bayesian reconstruction of single-photon 3d lidar data: Application to multispectral and underwater imaging.
\newblock {\em IEEE Transactions on Computational Imaging}, 7:961--974, 2021.

\bibitem{herholz2016product}
Sebastian Herholz, Oskar Elek, Ji{\v{r}}{\'\i} Vorba, Hendrik Lensch, and Jaroslav K{\v{r}}iv{\'a}nek.
\newblock Product importance sampling for light transport path guiding.
\newblock In {\em Computer Graphics Forum}, volume~35, pages 67--77. Wiley Online Library, 2016.

\bibitem{herholz2019volume}
Sebastian Herholz, Yangyang Zhao, Oskar Elek, Derek Nowrouzezahrai, Hendrik~PA Lensch, and Jaroslav K{\v{r}}iv{\'a}nek.
\newblock Volume path guiding based on zero-variance random walk theory.
\newblock {\em ACM Transactions on Graphics (TOG)}, 38(3):1--19, 2019.

\bibitem{hoogenboom2008zero}
J~Eduard Hoogenboom.
\newblock Zero-variance monte carlo schemes revisited.
\newblock {\em Nuclear science and Engineering}, 160(1):1--22, 2008.

\bibitem{jakob2010radiative}
Wenzel Jakob, Adam Arbree, Jonathan~T Moon, Kavita Bala, and Steve Marschner.
\newblock A radiative transfer framework for rendering materials with anisotropic structure.
\newblock In {\em ACM SIGGRAPH 2010 papers}, pages 1--13. 2010.

\bibitem{jarabo2012femto}
Adrian Jarabo.
\newblock Femto-photography: Visualizing light in motion.
\newblock {\em Universidad de Zaragoza}, 2012.

\bibitem{jarabo2018bidirectional}
Adrian Jarabo and Victor Arellano.
\newblock Bidirectional rendering of vector light transport.
\newblock In {\em Computer Graphics Forum}, volume~37, pages 96--105. Wiley Online Library, 2018.

\bibitem{jarabo2014framework}
Adrian Jarabo, Julio Marco, Adolfo Munoz, Raul Buisan, Wojciech Jarosz, and Diego Gutierrez.
\newblock A framework for transient rendering.
\newblock {\em ACM Transactions on Graphics (ToG)}, 33(6):1--10, 2014.

\bibitem{jarabo2017recent}
Adrian Jarabo, Belen Masia, Julio Marco, and Diego Gutierrez.
\newblock Recent advances in transient imaging: A computer graphics and vision perspective.
\newblock {\em Visual Informatics}, 1(1):65--79, 2017.

\bibitem{jarosz2011comprehensive}
Wojciech Jarosz, Derek Nowrouzezahrai, Iman Sadeghi, and Henrik~Wann Jensen.
\newblock A comprehensive theory of volumetric radiance estimation using photon points and beams.
\newblock {\em ACM transactions on graphics (TOG)}, 30(1):1--19, 2011.

\bibitem{jensen1996global}
Henrik~Wann Jensen.
\newblock Global illumination using photon maps.
\newblock In {\em Rendering Techniques’ 96: Proceedings of the Eurographics Workshop in Porto, Portugal, June 17--19, 1996 7}, pages 21--30. Springer, 1996.

\bibitem{jensen1998efficient}
Henrik~Wann Jensen and Per~H. Christensen.
\newblock Efficient simulation of light transport in scenes with participating media using photon maps.
\newblock In {\em Proceedings of the 25th Annual Conference on Computer Graphics and Interactive Techniques}, SIGGRAPH '98, page 311–320, New York, NY, USA, 1998. Association for Computing Machinery.

\bibitem{kijima2021time}
Daiki Kijima, Takahiro Kushida, Hiromu Kitajima, Kenichiro Tanaka, Hiroyuki Kubo, Takuya Funatomi, and Yasuhiro Mukaigawa.
\newblock Time-of-flight imaging in fog using multiple time-gated exposures.
\newblock {\em Optics Express}, 29(5):6453--6467, 2021.

\bibitem{kvrivanek2014unifying}
Jaroslav K{\v{r}}iv{\'a}nek, Iliyan Georgiev, Toshiya Hachisuka, Petr V{\'e}voda, Martin {\v{S}}ik, Derek Nowrouzezahrai, and Wojciech Jarosz.
\newblock Unifying points, beams, and paths in volumetric light transport simulation.
\newblock {\em ACM Transactions on Graphics (TOG)}, 33(4):1--13, 2014.

\bibitem{kulla2012importance}
Christopher Kulla and Marcos Fajardo.
\newblock Importance sampling techniques for path tracing in participating media.
\newblock In {\em Computer graphics forum}, volume~31, pages 1519--1528. Wiley Online Library, 2012.

\bibitem{lima2011improved}
Ivan~T Lima, Anshul Kalra, and Sherif~S Sherif.
\newblock Improved importance sampling for monte carlo simulation of time-domain optical coherence tomography.
\newblock {\em Biomedical optics express}, 2(5):1069--1081, 2011.

\bibitem{lin2021fast}
Daqi Lin, Chris Wyman, and Cem Yuksel.
\newblock Fast volume rendering with spatiotemporal reservoir resampling.
\newblock {\em ACM Transactions on Graphics (TOG)}, 40(6):1--18, 2021.

\bibitem{lister2012optical}
Tom Lister, Philip~A Wright, and Paul~H Chappell.
\newblock Optical properties of human skin.
\newblock {\em Journal of biomedical optics}, 17(9):090901--090901, 2012.

\bibitem{liu2022temporally}
Yang Liu, Shaojie Jiao, and Wojciech Jarosz.
\newblock Temporally sliced photon primitives for time-of-flight rendering.
\newblock {\em Computer Graphics Forum (Proceedings of EGSR)}, 41(4), July 2022.

\bibitem{marco2019progressive}
Julio Marco, Ib{\'o}n Guill{\'e}n, Wojciech Jarosz, Diego Gutierrez, and Adrian Jarabo.
\newblock Progressive transient photon beams.
\newblock In {\em Computer graphics forum}, volume~38, pages 19--30. Wiley Online Library, 2019.

\bibitem{marco2017deeptof}
Julio Marco, Quercus Hernandez, Adolfo Munoz, Yue Dong, Adrian Jarabo, Min~H Kim, Xin Tong, and Diego Gutierrez.
\newblock Deeptof: off-the-shelf real-time correction of multipath interference in time-of-flight imaging.
\newblock {\em ACM Transactions on Graphics (ToG)}, 36(6):1--12, 2017.

\bibitem{marco2017transient}
Julio Marco, Wojciech Jarosz, Diego Gutierrez, and Adrian Jarabo.
\newblock Transient photon beams.
\newblock In {\em ACM SIGGRAPH 2017 Posters}, pages 1--2. 2017.

\bibitem{novak2018monte}
Jan Nov{\'a}k, Iliyan Georgiev, Johannes Hanika, and Wojciech Jarosz.
\newblock Monte carlo methods for volumetric light transport simulation.
\newblock In {\em Computer graphics forum}, volume~37, pages 551--576. Wiley Online Library, 2018.

\bibitem{pan2019transient}
Xian Pan, Victor Arellano, and Adrian Jarabo.
\newblock Transient instant radiosity for efficient time-resolved global illumination.
\newblock {\em Computers \& Graphics}, 83:107--113, 2019.

\bibitem{pediredla2019ellipsoidal}
Adithya Pediredla, Ashok Veeraraghavan, and Ioannis Gkioulekas.
\newblock Ellipsoidal path connections for time-gated rendering.
\newblock {\em ACM Transactions on Graphics (TOG)}, 38(4):1--12, 2019.

\bibitem{periyasamy2016importance}
Vijitha Periyasamy and Manojit Pramanik.
\newblock Importance sampling-based monte carlo simulation of time-domain optical coherence tomography with embedded objects.
\newblock {\em Applied optics}, 55(11):2921--2929, 2016.

\bibitem{pharr2023physically}
Matt Pharr, Wenzel Jakob, and Greg Humphreys.
\newblock {\em Physically based rendering: From theory to implementation}.
\newblock MIT Press, 2023.

\bibitem{rapp2020seeing}
Joshua Rapp, Charles Saunders, Juli{\'a}n Tachella, John Murray-Bruce, Yoann Altmann, Jean-Yves Tourneret, Stephen McLaughlin, Robin~MA Dawson, Franco~NC Wong, and Vivek~K Goyal.
\newblock Seeing around corners with edge-resolved transient imaging.
\newblock {\em Nature communications}, 11(1):5929, 2020.

\bibitem{ren2008gradient}
Zhong Ren, Kun Zhou, Stephen Lin, and Baining Guo.
\newblock Gradient-based interpolation and sampling for real-time rendering of inhomogeneous, single-scattering media.
\newblock In {\em Computer Graphics Forum}, volume~27, pages 1945--1953. Wiley Online Library, 2008.

\bibitem{royo2023virtual}
Diego Royo, Talha Sultan, Adolfo Mu{\~n}oz, Khadijeh Masumnia-Bisheh, Eric Brandt, Diego Gutierrez, Andreas Velten, and Julio Marco.
\newblock Virtual mirrors: Non-line-of-sight imaging beyond the third bounce.
\newblock {\em ACM Transactions on Graphics (TOG)}, 42(4):1--15, 2023.

\bibitem{shem2020towards}
Kfir Shem-Tov, Sai~Praveen Bangaru, Anat Levin, and Ioannis Gkioulekas.
\newblock Towards reflectometry from interreflections.
\newblock In {\em 2020 IEEE International Conference on Computational Photography (ICCP)}, pages 1--12. IEEE, 2020.

\bibitem{talbot2005importance}
Justin~F Talbot.
\newblock {\em Importance resampling for global illumination}.
\newblock Brigham Young University, 2005.

\bibitem{veach1998robust}
Eric Veach.
\newblock {\em Robust Monte Carlo methods for light transport simulation}.
\newblock Stanford University, 1998.

\bibitem{velten2013femto}
Andreas Velten, Di~Wu, Adrian Jarabo, Belen Masia, Christopher Barsi, Chinmaya Joshi, Everett Lawson, Moungi Bawendi, Diego Gutierrez, and Ramesh Raskar.
\newblock Femto-photography: capturing and visualizing the propagation of light.
\newblock {\em ACM Transactions on Graphics (ToG)}, 32(4):1--8, 2013.

\bibitem{vorba2014line}
Ji{\v{r}}{\'\i} Vorba, Ond{\v{r}}ej Karl{\'\i}k, Martin {\v{S}}ik, Tobias Ritschel, and Jaroslav K{\v{r}}iv{\'a}nek.
\newblock On-line learning of parametric mixture models for light transport simulation.
\newblock {\em ACM Transactions on Graphics (TOG)}, 33(4):1--11, 2014.

\bibitem{walia2022gated2gated}
Amanpreet Walia, Stefanie Walz, Mario Bijelic, Fahim Mannan, Frank Julca-Aguilar, Michael Langer, Werner Ritter, and Felix Heide.
\newblock Gated2gated: Self-supervised depth estimation from gated images.
\newblock In {\em Proceedings of the IEEE/CVF Conference on Computer Vision and Pattern Recognition}, pages 2811--2821, 2022.

\bibitem{wu2018adaptive}
Rihui Wu, Adrian Jarabo, Jinli Suo, Feng Dai, Yongdong Zhang, Qionghai Dai, and Diego Gutierrez.
\newblock Adaptive polarization-difference transient imaging for depth estimation in scattering media.
\newblock {\em Optics Letters}, 43(6):1299--1302, 2018.

\bibitem{xin2019theory}
Shumian Xin, Sotiris Nousias, Kiriakos~N Kutulakos, Aswin~C Sankaranarayanan, Srinivasa~G Narasimhan, and Ioannis Gkioulekas.
\newblock A theory of fermat paths for non-line-of-sight shape reconstruction.
\newblock In {\em Proceedings of the IEEE/CVF conference on computer vision and pattern recognition}, pages 6800--6809, 2019.

\bibitem{yi2021differentiable}
Shinyoung Yi, Donggun Kim, Kiseok Choi, Adrian Jarabo, Diego Gutierrez, and Min~H Kim.
\newblock Differentiable transient rendering.
\newblock {\em ACM Transactions on Graphics (TOG)}, 40(6):1--11, 2021.

\bibitem{zhang2022first}
Tianyi Zhang, Mel~J White, Akshat Dave, Shahaboddin Ghajari, Ankit Raghuram, Alyosha~C Molnar, and Ashok Veeraraghavan.
\newblock First arrival differential lidar.
\newblock In {\em 2022 IEEE International Conference on Computational Photography (ICCP)}, pages 1--12. IEEE, 2022.

\bibitem{zickus2020fluorescence}
Vytautas Zickus, Ming-Lo Wu, Kazuhiro Morimoto, Valentin Kapitany, Areeba Fatima, Alex Turpin, Robert Insall, Jamie Whitelaw, Laura Machesky, Claudio Bruschini, et~al.
\newblock Fluorescence lifetime imaging with a megapixel spad camera and neural network lifetime estimation.
\newblock {\em Scientific Reports}, 10(1):20986, 2020.

\end{thebibliography}

%




\end{document}